\documentclass[12pt,a4paper]{article}
\pdfoutput=1
\usepackage{indentfirst}
\usepackage{graphicx}
\usepackage{subfigure}
\usepackage{threeparttable}
\usepackage{array}
\usepackage{multirow}
\usepackage{amsmath}
\usepackage{amssymb}
\usepackage{hyperref}
\hypersetup{pdfauthor={Ruifeng Yuan, Chengwen Zhong},pdftitle={A multi-prediction implicit scheme for steady state solutions of gas flow in all flow regimes}}

\begin{document}


\title{A multi-prediction implicit scheme for steady state solutions of gas flow in all flow regimes}
\renewcommand{\thefootnote}{\fnsymbol{footnote}}
\author{Ruifeng Yuan\footnotemark[1], Chengwen Zhong\footnotemark[1]}
\footnotetext[1]{National Key Laboratory of Science and Technology on Aerodynamic Design and Research, Northwestern Polytechnical University, Xi'an, Shaanxi 710072, China}
\footnotetext{\emph{Email addresses:} xyrfx@mail.nwpu.edu.cn (Ruifeng Yuan), zhongcw@nwpu.edu.cn (Chengwen Zhong)}
\date{May 15, 2019}
\maketitle


\rule[-5pt]{\textwidth}{0.5pt}
\begin{abstract}
An implicit multiscale method with multiple macroscopic prediction for steady state solutions of gas flow in all flow regimes is presented. The method is based on the finite volume discrete velocity method (DVM) framework. At the cell interface a multiscale flux with a construction similar to discrete unified gas-kinetic scheme (DUGKS) is adopted. The idea of the macroscopic variable prediction is further developed and a multiple prediction structure is formed. A prediction scheme is constructed to give a predicted macroscopic variable based on the macroscopic residual, and the convergence is accelerated greatly in the continuum flow regime. Test cases show the present method is one order of magnitude faster than the previous implicit multiscale scheme in the continuum flow regime.
~\\

\end{abstract}
\rule[5pt]{\textwidth}{0.5pt}


\section{Introduction}
Rarefied gas flow simulation is always a research hotspot of computational fluid dynamics (CFD). Recent years, multiscale gas-kinetic methods based on the discrete velocity method (DVM, \cite{Goldstein1989Investigations,Yang1995Rarefied,Mieussens2000Discretev,li2004Study,Titarev2007Conservative}) framework for nonequilibrium rarefied flow simulation have been developed, like the unified gas-kinetic scheme (UGKS) \cite{Xu2010A} by Xu and Huang, the discrete unified gas-kinetic scheme (DUGKS) \cite{guo2013discrete,guo2015discrete} by Guo et al. These multiscale methods overcome the time step and cell size restrictions of the original DVM method which requires time step and cell size of the order of mean collision time and mean free path, and thus have attracted more and more researchers' attention. It is worth pointing out that although UGKS and DUGKS can adopt time step and cell size comparable to the traditional macroscopic Navier-Stokes (NS) method, they still involve large amount of computation due to the curse of dimensionality. Hence, many researches on the acceleration of these multiscale methods have been carried out, including Mao et al.'s implicit UGKS \cite{Mao2015STUDY}, Zhu et al.'s prediction based implicit UGKS \cite{Zhu2016Implicit,zhu2019implicit}, Zhu et al.'s implicit multigrid UGKS algorithm \cite{Zhu2017Unified}, Yang et al.'s memory saving implicit multiscale scheme \cite{Yang2018An}, Pan et al.'s implicit DUGKS \cite{pan2019implicit}, etc. Following these previous works, it is quite valuable to further develop the fast algorithm for the multiscale method.

In this paper, a multiple prediction implicit multiscale method for steady state calculation of gas flow in all flow regimes is proposed. The idea of macroscopic prediction presented by Zhu et al.~\cite{Zhu2016Implicit} is further developed. A prediction solver is used to predict the macroscopic variable based on the macroscopic residual, and a multiple prediction procedure is constructed. The prediction solver is designed to ensure the accuracy of the predicted macroscopic variable in the continuum flow regime and the stability of the numerical system in all flow regimes, which makes the method very efficient in the continuum flow regime and stable in all flow regimes. Our test cases show that the present method is one order of magnitude faster than the previous implicit multiscale method in the continuum flow regime.

\section{Numerical method}\label{sec:numerical_method}
In this paper, the monatomic gas is considered and the governing equation is BGK-type equation \cite{bhatnagar1954model},
\begin{equation}\label{eq:bgk}
\frac{{\partial f}}{{\partial t}}{\rm{ + }}\vec u\cdot\frac{{\partial f}}{{\partial \vec x}} = \frac{{g - f}}{\tau },
\end{equation}
where $f$ is the gas particle velocity distribution function, $\vec u$ is the particle velocity, $\tau$ is the relaxation time calculated as $\tau  = \mu /p$ ($\mu$ and $p$ are the viscosity and pressure). $g$ is the equilibrium state which has a form of Maxwellian distribution,
\begin{equation}
g = \rho {\left( {\frac{\lambda }{\pi }} \right)^{\frac{3}{2}}}{e^{ - \lambda {{\vec c}^2}}},
\end{equation}
or if the Shakhov model \cite{shakhov1968generalization} is used
\begin{equation}\label{eq:eqstate}
g^* = \rho {\left( {\frac{\lambda }{\pi }} \right)^{\frac{3}{2}}}{e^{ - \lambda {{\vec c}^2}}}\left[ {1 + \frac{{4(1 - \Pr ){\lambda ^2}\vec q \cdot \vec c}}{{5\rho }}(2\lambda {{\vec c}^2} - 5)} \right],
\end{equation}
where $\vec c$ is the peculiar velocity $\vec c = \vec u - \vec U$ and $\vec U$ is the macroscopic gas velocity, $\vec q$ is the heat flux, $\lambda $ is a variable related to the temperature $T$ by $\lambda  = 1/(2RT)$. Pr is the Prandtl number and has a value of $2/3$ for monatomic gas. $f$ is related to the macroscopic variables by
\begin{equation}\label{eq:f_int_conserve}
\vec W = \int {\vec \psi fd\Xi},
\end{equation}
where $\vec W=(\rho,\rho\vec U,\rho E)^T$ is the vector of the macroscopic conservative variables, $\vec \psi$ is the vector of moments $\vec \psi  = {\left( {1,\vec u,\frac{1}{2}{{\vec u}^2}}\right)^T}$, $d\Xi  = du_xdu_ydu_z$ is the velocity space element. The stress tensor $\pmb{P}$ and the heat flux $\vec q$ can also be calculated by $f$ as
\begin{equation}\label{eq:f_int_stress}
    \pmb{P} = \int {\vec c\vec cfd\Xi },
\end{equation}
\begin{equation}\label{eq:f_int_qflux}
\vec q = \int {\frac{1}{2}\vec c{{\vec c}^2}fd\Xi }.
\end{equation}
Moreover, $f$ and $g$ obey the conservation law,
\begin{equation}\label{eq:int_conserve_law}
\int {\vec \psi (g - f)d\Xi }  = \vec 0.
\end{equation}
Adopting the integral form, the steady state of the governing equation Eq.~\ref{eq:bgk} is
\begin{equation}\label{eq:mic_fixedpoint}
\int\limits_{\partial \Omega } {\vec u \cdot \vec nfdA}  = \int\limits_\Omega  {\frac{{g - f}}{\tau }dV} ,
\end{equation}
where $\Omega$ is the control volume, $dV$ is the volume element, $dA$ is the surface area element and $\vec n$ is the outward normal unit vector. Take the moment of Eq.~\ref{eq:mic_fixedpoint} for $\vec \psi  = {\left( {1,\vec u,\frac{1}{2}{{\vec u}^2}}\right)^T}$, the corresponding macroscopic governing equation can be written as
\begin{equation}\label{eq:mac_fixedpoint}
\int\limits_{\partial \Omega } {\vec FdA}  = \vec 0,
\end{equation}
where the flux $\vec F$ has the relation with the distribution function $f$ by
\begin{equation}
\vec F = \int {\vec u \cdot \vec n\vec \psi fd\Xi }.
\end{equation}

\begin{figure}
\centering
\includegraphics[width=0.47\textwidth]{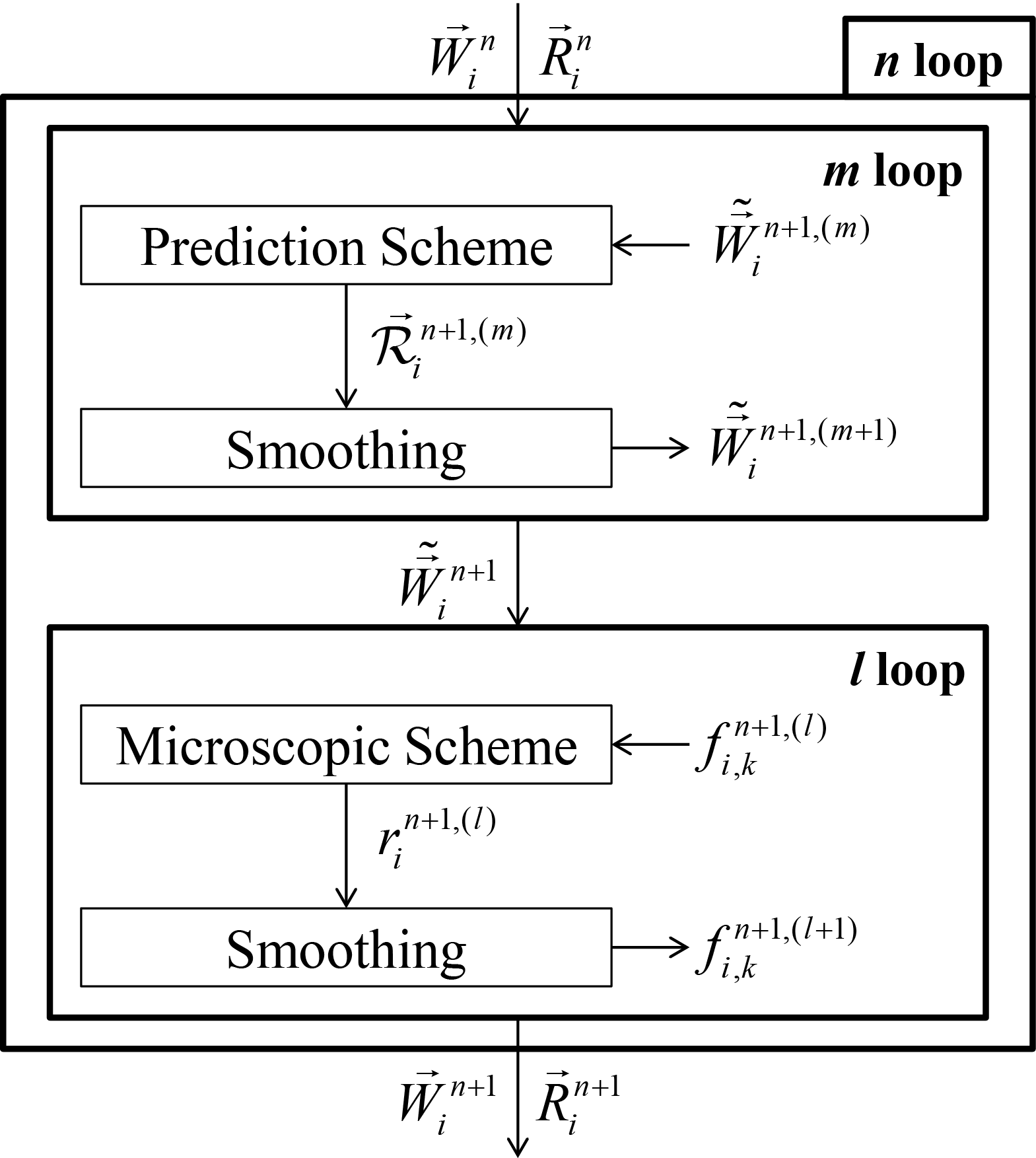}
\caption{\label{fig:general_frame}The general framework of the method.}
\end{figure}

This paper is about the numerical method of determining the steady state defined by Eq.~\ref{eq:mic_fixedpoint}. It is time-consuming to directly solve Eq.~\ref{eq:mic_fixedpoint} through a microscopic scheme involving discretization in both physical space and velocity space. The main idea of the present method is summarized as that, using the accurate but expensive scheme to calculate the residual of the system deviating from the steady state, then utilizing this residual and using the less accurate but efficient scheme to do the evolution. More distinctly, an accurate multiscale microscopic scheme based on the DVM framework is used to handle the microscopic numerical system with Eq.~\ref{eq:mic_fixedpoint}, and a fast prediction scheme is used to do the evolution of the macroscopic variables. The prediction scheme can be some kind of macroscopic scheme based on macroscopic variables or even a scheme based on the DVM framework but with less velocity points. The schematic of the general algorithmic framework for the present method is shown in Fig.~\ref{fig:general_frame}. The method consists of several loops in different layers. The outermost loop is denoted by $n$. One iteration of the $n$ loop includes a loop denoted by $m$ and a loop denoted by $l$, where the macroscopic variable $\vec W^{n}_{i}$ and the residual $\vec R^{n}_{i}$ are given as the input, the new $\vec W^{n+1}_{i}$ and $\vec R^{n+1}_{i}$ are the output. In the $m$ loop, the predicted macroscopic variable $\tilde{\vec W}^{n+1}_{i}$ is determined by the prediction scheme and by the numerical smoothing process. In the $l$ loop, the microscopic variable $f^{n+1}_{i,k}$ is calculated and the new $\vec W^{n+1}_{i}$ and $\vec R^{n+1}_{i}$ are obtained. The present method is a development of the prediction method of Zhu et al.~\cite{Zhu2016Implicit}, and has a structure similar to the multigrid method of Zhu et al.~\cite{Zhu2017Unified}, therefore we call it as ``multiple prediction method''. The method is detailed in following paragraphs.

\subsection{Construction of the $l$ loop}\label{sec:loop_l}
In the $l$ loop, residuals of the numerical system are evaluated through the microscopic scheme and the microscopic variables (the discrete distribution function) are updated through an implicit method (the numerical smoothing process). The microscopic scheme is very important because it determines the final steady state of the whole numerical system and thus determines the nature of the present numerical method.

The microscopic scheme is based on Eq.~\ref{eq:mic_fixedpoint}. Discretizing the physical space by finite volume method and discretizing the velocity space into discrete velocity points, the microscopic governing equation Eq.~\ref{eq:mic_fixedpoint} can be expressed as
\begin{equation}\label{eq:mic_fixedpoint_disc0}
\sum\limits_{j \in N\left( i \right)} {{A_{ij}}{{\vec u}_k} \cdot {{\vec n}_{ij}}f_{ij,k}^{}}  = {V_i}\frac{{g_{i,k}^{} - f_{i,k}^{}}}{{\tau _i^{}}},
\end{equation}
where the signs $i,k$ correspond to the discretizations in physical space and velocity space respectively. $j$ denotes the neighboring cell of cell $i$ and $N\left( i \right)$ is the set of all of the neighbors of $i$. Subscript $ij$ denotes the variable at the interface between cell $i$ and $j$. $A_{ij}$ is the interface area, ${\vec n_{ij}}$ is the outward normal unit vector of interface $ij$ relative to cell $i$, and $V_i$ is the volume of cell $i$. The $l$ loop aims to find the solution of Eq.~\ref{eq:mic_fixedpoint_disc0} with the input predicted variable $\tilde{\vec W}^{n+1}_{i}$, therefore Eq.~\ref{eq:mic_fixedpoint_disc0} can be written more exactly as
\begin{equation}\label{eq:mic_fixedpoint_disc}
\sum\limits_{j \in N\left( i \right)} {{A_{ij}}{{\vec u}_k} \cdot {{\vec n}_{ij}}f_{ij,k}^{n + 1}}  = {V_i}\frac{{\tilde g_{i,k}^{n + 1} - f_{i,k}^{n + 1}}}{{\tilde \tau _i^{n + 1}}},
\end{equation}
where the symbol $\sim$ denotes the predicted variables at the $(n+1)$th step. $\tilde g_{i,k}^{n + 1}$ and $\tilde \tau _i^{n + 1}$ can be directly calculated from the input variable $\tilde{\vec W}^{n+1}_{i}$. The distribution function $f_{ij,k}^{n + 1}$ at the interface $ij$ is very important to ensure the multiscale property of the scheme. In this paper, following the idea of DUGKS \cite{guo2013discrete,guo2015discrete}, the construction of $f_{ij,k}^{n + 1}$ in reference \cite{yuan2018conservative} is adopted, i.e.
\begin{equation}\label{eq:interfacef}
f_{ij,k}^{n + 1} = \frac{{\tilde \tau _{ij}^{n + 1}}}{{\tilde \tau _{ij}^{n + 1} + {h_{ij}}}}f\left( {{{\vec x}_{ij}} - {{\vec u}_k}{h_{ij}},0,{{\vec u}_k}} \right) + \frac{{{h_{ij}}}}{{\tilde \tau _{ij}^{n + 1} + {h_{ij}}}}\tilde g\left( {{{\vec x}_{ij}},0,{{\vec u}_k}} \right),
\end{equation}
where
\begin{equation}
f({\vec x_{ij}} - {\vec u_k}{h_{ij}},0,{\vec u_k}) = \left\{ {\begin{array}{*{20}{l}}
{f_{i,k}^{n + 1} + ({{\vec x}_{ij}} - {{\vec x}_i} - {{\vec u}_k}{h_{ij}})\nabla f_{i,k}^{n + 1}{\mkern 1mu} {\kern 1pt} {\mkern 1mu} {\mkern 1mu} {\kern 1pt} ,\;{\kern 1pt} \;{\kern 1pt} \;{\kern 1pt} {{\vec u}_k} \cdot {{\vec n}_{ij}} \ge 0,}\\
{f_{j,k}^{n + 1} + ({{\vec x}_{ij}} - {{\vec x}_j} - {{\vec u}_k}{h_{ij}})\nabla f_{j,k}^{n + 1}{\mkern 1mu} {\kern 1pt} {\mkern 1mu} {\kern 1pt} ,\;{\kern 1pt} \;{\kern 1pt} \;{\kern 1pt} {{\vec u}_k} \cdot {{\vec n}_{ij}} < 0.}
\end{array}} \right.
\end{equation}
In above equations, $\nabla f_{i,k}^{n+1}$ and $\nabla f_{j,k}^{n+1}$ can be obtained through the reconstruction of the distribution function data. $\tilde g\left( {{{\vec x}_{ij}},0,{{\vec u}_k}} \right)$ and ${\tilde \tau _{ij}^{n+1}}$ are calculated by the same way as the method of GKS \cite{xu2001gas} and they can be both calculated from the predicted macroscopic variable $\tilde {\vec W}^{n+1}_{i}$. For $\tilde g({\vec x_{ij}},0,{\vec u_k})$, it is determined by the interface macroscopic variables $\tilde {\vec W}^{n+1}_{ij}$, which can be calculated as
\begin{equation}
\tilde {\vec W}_{ij}^{n + 1} = \int_{\vec u\cdot{{\vec n}_{ij}} \ge 0} {\vec \psi \tilde g_{ij}^{\rm{l},n + 1}d\Xi  + } \int_{\vec u\cdot{{\vec n}_{ij}} < 0} {\vec \psi \tilde g_{ij}^{\rm{r},n + 1}d\Xi } ,
\end{equation}
where the superscripts $\rm{l}$ and $\rm{r}$ denote variables at the left and right sides of the interface, ${\tilde g_{ij}^{\rm{l},n + 1}}$ and ${\tilde g_{ij}^{\rm{r},n + 1}}$ can be determined after the spacial reconstruction of $\tilde {\vec W}^{n+1}_{i}$. For ${\tilde \tau _{ij}^{n+1}}$, it is calculated as
\begin{equation}
\tilde \tau _{ij}^{n + 1} = \frac{{\mu (\tilde {\vec W}_{ij}^{n + 1})}}{{p(\tilde {\vec W}_{ij}^{n + 1})}} + \frac{{\left| {p_{ij}^{{\rm{l}},n + 1} - p_{ij}^{{\rm{r}},n + 1}} \right|}}{{\left| {p_{ij}^{{\rm{l}},n + 1} + p_{ij}^{{\rm{r}},n + 1}} \right|}}{h_{ij}},
\end{equation}
where the pressure ${p_{ij}^{{\rm{l}},n + 1}}$, ${p_{ij}^{{\rm{r}},n + 1}}$ at two sides of the interface can be obtained from the reconstruction and the second term on the right is for artificial viscosity. $h_{ij}$ in above equations is calculated from the physical local time step
\begin{equation}
{h_{ij}} = \min ({h_i},{h_j}).
\end{equation}
The physical local time step $h_i$ for the cell $i$ is determined by the local CFL condition as
\begin{equation}
{h_i} = \frac{{{V_i}}}{{\mathop {\max }\limits_k \left( {\sum\limits_{j \in N(i)} {\left( {{{\vec u}_k} \cdot {{\vec n}_{ij}}{A_{ij}}{\rm{H}}[{{\vec u}_k} \cdot {{\vec n}_{ij}}]} \right)} } \right)}}{\rm{CFL}},
\end{equation}
where ${\rm{H}}[x]$ is the Heaviside function defined as
\begin{equation}
{\rm{H}}[x] = \left\{ \begin{array}{l}
0,\quad x < 0,\\
1,\quad x \ge 0.
\end{array} \right.
\end{equation}
For more details about the construction of the interface distribution function $f_{ij,k}^{n + 1}$ please refer to reference \cite{yuan2018conservative}.

Eq.~\ref{eq:mic_fixedpoint_disc} is solved by iterations. The microscopic residual $r_{i,k}^{n + 1,(l)}$ at the $l$th iteration can be defined as
\begin{equation}\label{eq:mic_residual}
r_{i,k}^{n + 1,(l)} = \frac{{\tilde g_{i,k}^{n + 1} - f_{i,k}^{n + 1,(l)}}}{{\tilde \tau _i^{n + 1}}} - \frac{1}{V_i}\sum\limits_{j \in N\left( i \right)} {{A_{ij}}{{\vec u}_k} \cdot {{\vec n}_{ij}}f_{ij,k}^{n + 1,(l)}}.
\end{equation}
According to the previous descriptions, $r_{i,k}^{n + 1,(l)}$ can be calculated from $f_{i,k}^{n + 1,(l)}$ and $\tilde {\vec W}^{n+1}_{i}$ through the spatial data reconstruction. The increment equation to get the microscopic variable $f_{i,k}^{n + 1,(l+1)}$ at the iteration $l+1$ is constructed by backward Euler method,
\begin{equation}\label{eq:mic_iter_rsd}
r_{i,k}^{n + 1,(l)} + \Delta r_{i,k}^{n + 1,(l + 1)} = \frac{1}{{\Delta \xi _{i,k}^{n + 1,(l + 1)}}}\Delta f_{i,k}^{n + 1,(l + 1)},
\end{equation}
where ${\Delta \xi _{i,k}^{n + 1,(l + 1)}}$ is the pseudo time step and ${\Delta \xi _{i,k}^{n + 1,(l + 1)}}$ is always set to be $\infty$ in the present study. Combined with the residual expression Eq.~\ref{eq:mic_residual}, Eq.~\ref{eq:mic_iter_rsd} can be written as
\begin{equation}\label{eq:mic_iter}
\left( {\frac{1}{{\Delta \xi _{i,k}^{n + 1,(l + 1)}}} + \frac{1}{{\tilde \tau _i^{n + 1}}}} \right)\Delta f_{i,k}^{n + 1,(l + 1)} = r_{i,k}^{n + 1,(l)} - \frac{1}{V_i}\sum\limits_{j \in N(i)} {{A_{ij}}{{\vec u}_k} \cdot {{\vec n}_{ij}}\Delta f_{ij,k}^{n + 1,(l + 1)}}.
\end{equation}
For the increment of the interface distribution function ${\Delta f_{ij,k}^{n + 1,(l + 1)}}$, it is simply handled by a modified upwind scheme
\begin{equation}\label{eq:mic_itff_jne0}
\Delta f_{ij,k}^{n + 1,(l + 1)} = \left\{ \begin{array}{l}
\frac{{\tilde \tau _{ij}^{n + 1}}}{{\tilde \tau _{ij}^{n + 1} + {h_{ij}}}}\Delta f_{i,k}^{n + 1,(l + 1)},\quad {{\vec u}_k} \cdot {{\vec n}_{ij}} \ge 0\\
\frac{{\tilde \tau _{ij}^{n + 1}}}{{\tilde \tau _{ij}^{n + 1} + {h_{ij}}}}\Delta f_{j,k}^{n + 1,(l + 1)},\quad {{\vec u}_k} \cdot {{\vec n}_{ij}} < 0
\end{array} \right. ,
\end{equation}
where the coefficient ${\frac{{\tilde \tau _{ij}^{n + 1}}}{{\tilde \tau _{ij}^{n + 1} + {h_{ij}}}}}$ is the corresponding coefficient multiplied by $f\left( {{{\vec x}_{ij}} - {{\vec u}_k}{h_{ij}},0,{{\vec u}_k}} \right)$ in Eq.~\ref{eq:interfacef}. This coefficient is multiplied because during the whole $l$ loop the term $\tilde g\left( {{{\vec x}_{ij}},0,{{\vec u}_k}} \right)$ in Eq.~\ref{eq:interfacef} is calculated by the predicted macroscopic variable $\tilde {\vec W}^{n+1}_{i}$ and therefore is an invariant, so the variation of the microscopic variable $f_{i,k}^{n + 1,(l + 1)}$ only influences the term $f\left( {{{\vec x}_{ij}} - {{\vec u}_k}{h_{ij}},0,{{\vec u}_k}} \right)$, which is multiplied by the coefficient ${\frac{{\tilde \tau _{ij}^{n + 1}}}{{\tilde \tau _{ij}^{n + 1} + {h_{ij}}}}}$. It is worth noting that in an actual implementation of the method, the interface distribution function $f_{ij,k}^n$ at the $n$th step may be taken as the initial value $f_{ij,k}^{n+1,(0)}$ at $l=0$ for the step $n+1$ to reduce computation cost, in this situation the variation $\Delta f_{i,k}^{n + 1,(1)}$ should also account for the variation of $\tilde g\left( {{{\vec x}_{ij}},0,{{\vec u}_k}} \right)$, and the coefficient ${\frac{{\tilde \tau _{ij}^{n + 1}}}{{\tilde \tau _{ij}^{n + 1} + {h_{ij}}}}}$ shouldn't be multiplied at the first iteration of the $l$ loop, i.e.
\begin{equation}\label{eq:mic_itff_je0}
\Delta f_{ij,k}^{n + 1,(1)} = \left\{ {\begin{array}{*{20}{l}}
{\Delta f_{i,k}^{n + 1,(1)},\quad {{\vec u}_k} \cdot {{\vec n}_{ij}} \ge 0}\\
{\Delta f_{j,k}^{n + 1,(1)},\quad {{\vec u}_k} \cdot {{\vec n}_{ij}} < 0}
\end{array}} \right..
\end{equation}
In this situation, after the first iteration of the $l$ loop, the interface distribution function $f_{ij,k}^{n + 1,(l>0)}$ will be calculated with the newly predicted $\tilde {\vec W}^{n+1}_{i}$ and Eq.~\ref{eq:mic_itff_jne0} is used to handle ${\Delta f_{ij,k}^{n + 1,(l + 1)}}$ again. Without loss of generality, substituting Eq.~\ref{eq:mic_itff_jne0} into Eq.~\ref{eq:mic_iter} will yield
\begin{equation}\label{eq:mic_update}
\begin{aligned}
& \left( {\frac{1}{{\Delta \xi _{i,k}^{n + 1,(l + 1)}}} + \frac{1}{{\tilde \tau _i^{n + 1}}} + \frac{1}{V_i}\sum\limits_{j \in N_k^ + (i)} {\frac{{\tilde \tau _{ij}^{n + 1}}}{{\tilde \tau _{ij}^{n + 1} + {h_{ij}}}}{A_{ij}}{{\vec u}_k}\cdot{{\vec n}_{ij}}} } \right)\Delta f_{i,k}^{n + 1,(l + 1)}\\
 =  & r_{i,k}^{n + 1,(l)} - \frac{1}{V_i}\sum\limits_{j \in N_k^ - (i)} {\frac{{\tilde \tau _{ij}^{n + 1}}}{{\tilde \tau _{ij}^{n + 1} + {h_{ij}}}}{A_{ij}}{{\vec u}_k}\cdot{{\vec n}_{ij}}\Delta f_{j,k}^{n + 1,(l + 1)}} ,
\end{aligned}
\end{equation}
where $ N_k^ + (i)$ is the set of $i$'s neighboring cells satisfying ${\vec u_k} \cdot {\vec n_{ij}} \ge 0$ while for $ N_k^ - (i)$ it satisfies ${\vec u_k} \cdot {\vec n_{ij}} < 0$. For simplicity, Eq.~\ref{eq:mic_update} is solved by the Symmetric Gauss-Seidel (SGS) method, or also known as the Point Relaxation Symmetric Gauss-Seidel (PRSGS) method \cite{Rogers1995Comparison,Yuan2002Comparison}. In each time of the SGS iteration, a forward sweep from the first to the last cell and a backward sweep from the last to the first cell are implemented, during which the data of a cell is always updated by the latest data of its adjacent cells through Eq.~\ref{eq:mic_update}. Such a SGS iteration procedure is totally matrix-free and easy to implement.

After several times of SGS iterations for solving Eq.~\ref{eq:mic_update}, an evaluation of $f_{i,k}^{n + 1,(l + 1)}$ with a certain precision can be obtained. Then the residual $r_{i,k}^{n + 1,(l+1)}$ at the $(l+1)$th iteration of the $l$ loop can be computed from $f_{i,k}^{n + 1,(l + 1)}$ and $\tilde {\vec W}^{n+1}_{i}$, and a new turn of the $l$ loop will be performed. After several iterations of the $l$ loop, an evaluation of $f_{i,k}^{n + 1}$ with a certain precision can be obtained, and the interface distribution function $f_{ij,k}^{n + 1}$ can be calculated by Eq.~\ref{eq:interfacef}. Then the macroscopic numerical flux ${\vec F_{ij}^{n + 1}}$ at the interface can be got by numerical integral in the discrete velocity space
\begin{equation}\label{eq:f_inc_disc_flux}
\vec F_{ij}^{n + 1} = \sum\limits_k {{{\vec \psi }_k}{{\vec u}_k} \cdot {{\vec n}_{ij}}f_{ij,k}^{n + 1}\Delta {\Xi _k}} ,
\end{equation}
and the macroscopic residual $\vec R_i^{n + 1}$ defined by the macroscopic governing equation Eq.~\ref{eq:mac_fixedpoint} at the $(n+1)$th step can be calculated from the flux by
\begin{equation}\label{eq:mac_fluxrsd}
\vec R_i^{n + 1} =  - \frac{1}{{{V_i}}}\sum\limits_{j \in N(i)} {{A_{ij}}\vec F_{ij}^{n + 1}}.
\end{equation}
Note that in the $l$ loop we solve the microscopic system Eq.~\ref{eq:mic_fixedpoint_disc} which is under the condition of the predicted variable $\tilde {\vec W}^{n+1}_{i}$, so $\vec R_i^{n + 1}$ is not zero even if the microscopic system is solved sufficiently accurately. Finally, the macroscopic variable ${\vec W}^{n+1}_{i}$ is calculated by numerical integral as
\begin{equation}\label{eq:f_inc_disc_mac}
\vec W_i^{n + 1} = \sum\limits_k {{{\vec \psi }_k}f_{i,k}^{n + 1}\Delta {\Xi _k}}  + \tilde {\vec W}_i^{n + 1} - \sum\limits_k {{{\vec \psi }_k}\tilde g_{i,k}^{n + 1}\Delta {\Xi _k}} ,
\end{equation}
where the term $\tilde {\vec W}_i^{n + 1} - \sum\limits_k {{{\vec \psi }_k}\tilde g_{i,k}^{n + 1}\Delta {\Xi _k}} $ is the integral error compensation term to make the scheme conservative, more details about this term please refer to reference \cite{yuan2018conservative}.

The iteration of the $l$ loop is similar to the numerical smoothing process in multigrid method \cite{Zhu2017Unified}. The computation procedure of the $l$ loop is listed as follows:
\begin{description}
    \item[Step 1.] Set the initial value $f_{i,k}^{n + 1,(0)}=f_{i,k}^{n}$.
    \item[Step 2.] Calculate the interface distribution function $f_{ij,k}^{n + 1,(l)}$  by Eq.~\ref{eq:interfacef} from $f_{i,k}^{n + 1,(l)}$ and $\tilde {\vec W}_i^{n + 1}$ through data spatial reconstruction. Calculate the microscopic residual $r_{i,k}^{n + 1,(l)}$ by Eq.~\ref{eq:mic_residual}.
    \item[Step 3.] Make judgement: if the residual $r_{i,k}^{n + 1,(l)}$ meets the convergence criterion, or if the iteration number of the $l$ loop meets the maximum limit, break out of the $l$ loop and go to Step 5.
    \item[Step 4.] Solve Eq.~\ref{eq:mic_update} by several times of SGS iterations, obtain $f_{i,k}^{n + 1,(l+1)}$, and go to Step 2.
    \item[Step 5.] By Eq.~\ref{eq:f_inc_disc_flux}, Eq.~\ref{eq:mac_fluxrsd} and Eq.~\ref{eq:f_inc_disc_mac}, do numerical integral in the velocity space to get ${\vec W}_{i}^{n+1}$ and $\vec R_i^{n + 1}$ for the step $n+1$.
\end{description}

\subsection{Construction of the $m$ loop}\label{sec:multiscaleflux}
In the $m$ loop, based on the macroscopic variable ${\vec W}_{i}^{n}$ and the macroscopic residual $\vec R_i^{n}$ at the $n$th step, a reasonable estimation for the macroscopic variable $\tilde {\vec W}_i^{n + 1}$ is obtained through a fast prediction scheme to accelerate convergence. Theoretically speaking, the prediction scheme can be either a macroscopic scheme based on macroscopic variables or a microscopic scheme based on the DVM framework but with less velocity points. In this paper, a macroscopic scheme is designed to do the prediction. The process of the $m$ loop has certain similarity to the coarse grid correction in the multigrid method \cite{Zhu2017Unified}.

\subsubsection{Framework}
The macroscopic residual has the form of Eq.~\ref{eq:mac_fluxrsd}. To reduce the residual, a prediction equation is constructed by backward Euler formula
\begin{equation}\label{eq:mac_prediction}
\frac{1}{{\Delta {t_i^{n + 1}}}}\left( {\tilde {\vec W}_i^{n + 1} - \vec W_i^n} \right) = \vec R_i^n + \Delta \tilde {\vec R}_i^{n + 1}.
\end{equation}
$\Delta {t_i^{n + 1}}$ is the local prediction time step, the purpose of this time step is to constrain the marching time depth of the prediction process to make the scheme stable in the extreme case. The predicted residual increment $\Delta \tilde {\vec R}_i^{n + 1}$ is calculated by
\begin{equation}\label{eq:mac_prediction_rsd}
\Delta \tilde {\vec R}_i^{n + 1} =  - \frac{1}{{{V_i}}}\sum\limits_{j \in N(i)} {{A_{ij}}\tilde {\vec {\mathcal{F}}}_{ij}^{n + 1}}  + \frac{1}{{{V_i}}}\sum\limits_{j \in N(i)} {{A_{ij}}\vec {\mathcal{F}}_{ij}^n},
\end{equation}
where $\vec {\mathcal{F}}_{ij}^n$ and $\tilde {\vec {\mathcal{F}}}_{ij}^{n + 1}$ are fluxes calculated by the prediction solver from $\vec W_{i}^n$ and the predicted $\tilde {\vec W}_{i}^{n + 1}$ with data reconstruction. This prediction solver is well-designed to balance between accuracy and stability, and will be presented later in the next section.

The aim of the $m$ loop is to solve Eq.~\ref{eq:mac_prediction} and give an estimation for $\tilde {\vec W}_{i}^{n + 1}$ with a certain precision. Like what we do in the $l$ loop, Eq.~\ref{eq:mac_prediction} is also solved by iterations. The residual $\vec {\mathcal{R}}_i^{n+1,(m)}$ at the $m$th iteration can be defined by Eq.~\ref{eq:mac_prediction} and expressed as
\begin{equation}\label{eq:mac_iter_rsd}
\begin{aligned}
\vec {\mathcal{R}}_i^{n+1,(m)} = & - \frac{1}{{{V_i}}}\sum\limits_{j \in N(i)} {{A_{ij}}\tilde {\vec {\mathcal{F}}}_{ij}^{n + 1,(m)}}  + \frac{1}{{{V_i}}}\sum\limits_{j \in N(i)} {{A_{ij}}\vec {\mathcal{F}}_{ij}^n}  + \vec R_i^n \\
 & - \frac{1}{{\Delta t_i^{n + 1}}}\left( {\tilde {\vec W}_i^{n + 1,(m)} - \vec W_i^n} \right),
\end{aligned}
\end{equation}
and the corresponding increment equation for $\tilde {\vec W}_i^{n + 1,(m+1)}$ is
\begin{equation}\label{eq:mac_iter}
\vec {\mathcal{R}}_i^{n + 1,(m)} + \Delta \vec {\mathcal{R}}_i^{n + 1,(m + 1)} = \frac{1}{{\Delta \eta _i^{n + 1,(m + 1)}}}\Delta \tilde {\vec W}_i^{n + 1,(m + 1)},
\end{equation}
where ${\Delta \eta _i^{n + 1,(m + 1)}}$ is the pseudo time step. Considering Eq.~\ref{eq:mac_iter_rsd}, the increment of the residual $\Delta \vec {\mathcal{R}}_i^{n + 1,(m + 1)}$ can be expressed as
\begin{equation}\label{eq:mac_iter_rsd_inc}
\Delta \vec {\mathcal{R}}_i^{n + 1,(m + 1)} = - \frac{1}{{\Delta t_i^{n + 1}}}\Delta \tilde {\vec W}_i^{n + 1,(m + 1)} - \frac{1}{{{V_i}}}\sum\limits_{j \in N(i)} {{A_{ij}}\Delta \tilde {\vec {\mathcal{F}}}_{ij}^{n + 1,(m+1)}},
\end{equation}
the variation of the flux $\Delta \tilde {\vec {\mathcal{F}}}_{ij}^{n + 1,(m+1)}$ is further approximated by
\begin{equation}\label{eq:mac_iter_flux_inc}
\Delta \tilde {\vec {\mathcal{F}}}_{ij}^{n + 1,(m+1)} =  {\vec {\mathsf{F}}}_{ij}^{n + 1,(m+1)} - {\vec {\mathsf{F}}}_{ij}^{n + 1,(m)},
\end{equation}
where $\vec {\mathsf{F}}_{ij}$ has the form \cite{luo1998fast} of the well-known Roe's flux function
\begin{equation}\label{eq:mac_iter_flux_roe}
\vec {\mathsf{F}}_{ij} = \frac{1}{2}\left( {{{\vec {\mathbb{F}}}_{ij}}({\vec W}_i) + {{\vec {\mathbb{F}}}_{ij}}({\vec W}_j) + {\mathfrak{r}_{ij}}{{\vec W}_i} - {\mathfrak{r}_{ij}}{{\vec W}_j}} \right).
\end{equation}
Here ${\vec {\mathbb{F}}_{ij}}(\vec W)$ is the Euler flux
\begin{equation}
{\vec {\mathbb{F}}_{ij}}(\vec W) = \left( \begin{array}{c}
\rho \vec U \cdot {{\vec n}_{ij}}\\
\rho {U_x}\vec U \cdot {{\vec n}_{ij}} + {n_{ij,x}}p\\
\rho {U_y}\vec U \cdot {{\vec n}_{ij}} + {n_{ij,y}}p\\
\rho {U_z}\vec U \cdot {{\vec n}_{ij}} + {n_{ij,z}}p\\
(\rho E + p)\vec U \cdot {{\vec n}_{ij}}
\end{array} \right),
\end{equation}
and ${\mathfrak{r}_{ij}}$ is
\begin{equation}
{\mathfrak{r}_{ij}} = \left| {{{\vec U}_{ij}} \cdot {{\vec n}_{ij}}} \right| + {a_{ij}} + 2\frac{{{\mu _{ij}}}}{{{\rho _{ij}}\Delta {x_{ij}}}},
\end{equation}
where $a_{ij}$ is the acoustic speed at the interface and $\Delta {x_{ij}}$ is the distance between cell center $i$ and $j$. Substitute Eq.~\ref{eq:mac_iter_rsd_inc}, Eq.~\ref{eq:mac_iter_flux_inc} and Eq.~\ref{eq:mac_iter_flux_roe} into Eq.~\ref{eq:mac_iter}, approximate ${{\mathfrak{r}}_{ij}^{n+1,(m+1)}}$ by ${\mathfrak{r}_{ij}^{n+1,(m)}}$, and note that $\sum\limits_{j \in N(i)} {{A_{ij}}{\vec{\mathbb{F}}_{ij}}({{\vec W}_i})}  = \vec 0$ holds, we can get
\begin{equation}\label{eq:mac_update}
\begin{aligned}
& \left( {\frac{1}{{\Delta t_i^{n + 1}}} + \frac{1}{{\Delta \eta _i^{n + 1,(m + 1)}}} + \frac{1}{{2{V_i}}}\sum\limits_{j \in N(i)} {{\mathfrak{r}}_{ij}^{n + 1,(m)}{A_{ij}}} } \right)\Delta \tilde {\vec W}_i^{n + 1,(m + 1)} \\
=  & \vec {\mathcal {R}}_i^{n + 1,(m)} + \frac{1}{{2{V_i}}}\sum\limits_{j \in N(i)} {{\mathfrak{r}}_{ij}^{n + 1,(m)}{A_{ij}}\Delta \tilde {\vec W}_j^{n + 1,(m + 1)}} \\
& - \frac{1}{{2{V_i}}}\sum\limits_{j \in N(i)} {{A_{ij}}\left( {{{\vec {\mathbb{F}}}_{ij}}(\tilde {\vec W}_j^{n + 1,(m + 1)}) - {{\vec {\mathbb{F}}}_{ij}}(\tilde {\vec W}_j^{n + 1,(m)})} \right)} .
\end{aligned}
\end{equation}
Eq.~\ref{eq:mac_update} is solved by several times' SGS iterations. An estimation of $\tilde {\vec W}_i^{n + 1,(m + 1)}$ with a certain precision can be obtained from Eq.~\ref{eq:mac_update}, then ${\mathfrak{r}}_{ij}^{n + 1,(m+1)}$ and the residual $\vec {\mathcal {R}}_i^{n + 1,(m+1)}$ at the $(m+1)$th iteration of the $m$ loop can be calculated. After several turns of the $m$ loop, the predicted macroscopic variable $\tilde {\vec W}_i^{n + 1}$ with a certain precision can be determined.

In fact, utilizing ${\vec W}_i^{n}$ and $\tilde {\vec W}_i^{n + 1}$, a prediction for the microscopic variable ${\tilde f}_{i,k}^{n + 1}$ can also be obtained to accelerate the convergence of the microscopic numerical system (i.e. the $l$ loop). The increment of the distribution function $\Delta {\tilde f}_{i,k}^{n + 1}$  can be calculated from the Chapman-Enskog expansions \cite{chapman1990mathematical} based on macroscopic variables ${\vec W}_i^{n}$ and $\tilde {\vec W}_i^{n + 1}$. This strategy will increase the complexity of the algorithm and thus is not adopted in the present method.

Likewise, as one can see, the process of the $m$ loop is similar to the numerical smoothing process in multigrid method \cite{Zhu2017Unified}. The computation procedure of the $m$ loop is listed as follows:
\begin{description}
    \item[Step 1.] Set the initial value $\tilde {\vec W}_i^{n + 1,(0)}={\vec W}_i^n$.
    \item[Step 2.] Calculate the residual $\vec {\mathcal {R}}_i^{n + 1,(m)}$ by Eq.~\ref{eq:mac_iter_rsd} from $\vec R_{i}^n$, $\vec W_{i}^n$ and $\tilde {\vec W}_i^{n + 1,(m)}$ (data reconstruction is implemented).
    \item[Step 3.] Make judgement: if the residual $\vec {\mathcal {R}}_i^{n + 1,(m)}$ meets the convergence criterion, or if the iteration number of the $m$ loop meets the maximum limit, break out of the $m$ loop and the predicted macroscopic variable $\tilde {\vec W}_{i}^{n+1}$ is determined.
    \item[Step 4.] Solve Eq.~\ref{eq:mac_update} by several times of SGS iterations, obtain $\tilde {\vec W}_i^{n + 1,(m+1)}$, and go to Step 2.
\end{description}

\subsubsection{Prediction solver}
The prediction solver used to calculate the fluxes $\vec {\mathcal{F}}_{ij}^n$ and $\tilde {\vec {\mathcal{F}}}_{ij}^{n + 1}$ in Eq.~\ref{eq:mac_prediction_rsd} requires careful design. For the continuum flow, the prediction solver should be as accurate as a traditional NS solver. For the rarefied flow, it's unrealistic for a fast solver based on macroscopic variables to provide a very precise flux, but the solver should be stable so that the present method can be applied to all flow regimes. Thus, there are two principles for the prediction solver: accurate in the continuum flow regime, stable in all flow regimes.

We start constructing the solver from the view of gas kinetic theory. Based on the famous Chapman-Enskog expansion \cite{chapman1990mathematical}, the distribution function $f$ obtained from the BGK equation Eq.~\ref{eq:bgk} to the first order of $\tau$ is
\begin{equation}\label{eq:bgk_ce1}
f = g - \tau (\frac{{\partial g}}{{\partial t}} + \vec u \cdot \frac{{\partial g}}{{\partial \vec x}}).
\end{equation}
Suppose there is an interface in $x$ direction. If the interface distribution function has the form of Eq.~\ref{eq:bgk_ce1}, take moments of $u_x\vec \psi$ to Eq.~\ref{eq:bgk_ce1} and ignore second (and higher) order terms of $\tau$, we can get the NS flux \cite{xu2001gas,Xu2015Direct}, where the term $g$ corresponds to the Euler flux and terms with $\tau$ (i.e. terms except $g$) correspond to viscous terms in the NS flux. Flux directly calculated from Eq.~\ref{eq:bgk_ce1} will lead to divergence in many cases, and we introduce some modifications below.

The Euler flux often causes stability issue. Inspiring by gas-kinetic scheme (GKS) or also known as BGK-NS scheme \cite{xu2001gas}, we replace it by a weighting of Euler flux and the flux of kinetic flux vector splitting (KFVS) \cite{mandal1994kinetic}. That is, we replace the term $g$ in Eq.~\ref{eq:bgk_ce1} and the interface distribution function is expressed as
\begin{equation}\label{eq:mac_interfacef_0}
f = \frac{\tau' }{{\tau'  + h}}{g^{\rm{lr}}} + \frac{h}{{\tau'  + h}}{g} - \tau (\frac{{\partial g}}{{\partial t}} + \vec u \cdot \frac{{\partial g}}{{\partial \vec x}}),
\end{equation}
where $g^{\rm{lr}}$ is
\begin{equation}
{g^{{\rm{lr}}}} = \left\{ \begin{array}{l}
{g^{\rm{l}}},u_x \ge 0\\
{g^{\rm{r}}},u_x < 0
\end{array} \right.
\end{equation}
which is determined by the reconstructed macroscopic variables on the two side of the interface. The interface macroscopic variable $\vec W$ is calculated as
\begin{equation}
\vec W = \int_{u_x \ge 0} {\vec \psi {g^{\rm{l}}}d\Xi }  + \int_{u_x < 0} {\vec \psi {g^{\rm{r}}}d\Xi } ,
\end{equation}
and $g$ is obtained from $\vec W$. The weight factors $\tau' /(\tau'  + h)$ and $h /(\tau'  + h)$ share the same forms as those in Eq.~\ref{eq:interfacef} (for how these weight factors are constructed please refer to \cite{yuan2018conservative}), and $\tau'$ is calculated by
\begin{equation}
\tau ' = \tau  + {\tau _{{\rm{artificial}}}} = \frac{\mu }{p} + \frac{{\left| {{p^{\rm{l}}} -{p^{\rm{r}}}} \right|}}{{\left| {{p^{\rm{l}}} + {p^{\rm{r}}}} \right|}}h,
\end{equation}
where ${\tau _{{\rm{artificial}}}}$ is for artificial viscosity. $h$ is the local CFL time step and is equal to $h_{ij}$ in Eq.~\ref{eq:interfacef}. Eq.~\ref{eq:mac_interfacef_0} has a form similar to the interface distribution function of GKS \cite{xu2001gas}, except that the viscous term is not upwind split and the weight factor is constructed following the thought of DUGKS \cite{guo2013discrete,guo2015discrete}. Because the KFVS scheme is very robust, the flux obtained from Eq.~\ref{eq:mac_interfacef_0} makes the numerical system more stable than directly using Eq.~\ref{eq:bgk_ce1}. In the continuum flow regime, $h\gg\tau$, if the flow is continuous the term ${\tau _{{\rm{artificial}}}}$ for artificial viscosity will be negligible and Eq.~\ref{eq:mac_interfacef_0} will recover the NS flux, while if the flow is discontinuous the term ${\tau _{{\rm{artificial}}}}$ will be activated and Eq.~\ref{eq:mac_interfacef_0} will work as a stable KFVS solver. In the rarefied flow simulation, $\tau > h$ and the inviscid part of Eq.~\ref{eq:mac_interfacef_0} generally provides a KFVS flux, which can increase the stability of the scheme.

The flux obtained from Eq.~\ref{eq:mac_interfacef_0} works well in the continuum flow regime. However, in the case of large Kn number, the numerical system based on Eq.~\ref{eq:mac_interfacef_0} is very stiff due to the large NS-type linear viscous term and the scheme is easy to blow up. Therefore, we multiply the viscous term by a limiting factor ${\mathfrak{q}}(\kappa)$ and Eq.~\ref{eq:mac_interfacef_0} is transformed into
\begin{equation}\label{eq:mac_interfacef}
f = \frac{\tau' }{{\tau'  + h}}{g^{\rm{lr}}} + \frac{h}{{\tau'  + h}}{g} - {\mathfrak{q}}(\kappa)\tau (\frac{{\partial g}}{{\partial t}} + \vec u \cdot \frac{{\partial g}}{{\partial \vec x}}).
\end{equation}
Here we emphasize that the limiting factor ${\mathfrak{q}}(\kappa)$ aims not to accurately calculate the flux, but to increase the stability in the case of large Kn number. One can view it as an empirical parameter. The limiting factor ${\mathfrak{q}}(\kappa)$ is constructed considering the form of nonlinear coupled constitutive relations (NCCR) \cite{xiao2014computational,liu2019Anextended}, and is expressed as
\begin{equation}\label{eq:mac_interfacef_nccrlim}
{\mathfrak{q}}(\kappa) = \frac{\kappa}{\sinh (\kappa)},
\end{equation}
which has $\mathop {\lim }\limits_{\kappa  \to 0} {\mathfrak{q}}(\kappa ) = 1$ and $\mathop {\lim }\limits_{\kappa  \to +\infty } {\mathfrak{q}}(\kappa ) = 0$. $\kappa$ is related to the viscous term and calculated as
\begin{equation}\label{eq:mac_interfacef_nccrkap}
\kappa  = \ln \left( {2\frac{{{\pi ^{\frac{1}{4}}}}}{{\sqrt {2\beta } }}\sqrt {\frac{{\Pr {{\left| {k\nabla T} \right|}^2}}}{{{C_p}T{p^2}}} + \frac{{{{\left| {2\mu {S_{ij}}} \right|}^2}}}{{2{p^2}}}}  + 1} \right),
\end{equation}
where $-k\nabla T$ and $2\mu {S_{ij}}$ correspond to the heat flux and stress in NS equation, $C_p$ is the specific heat at constant pressure. $\beta$ is a molecular model coefficient \cite{liu2019Anextended} involved in the variable soft sphere (VSS) model \cite{Koura1991Variable,Koura1992Variable} and is calculated as
\begin{equation}
\beta  = \frac{{5(\alpha  + 1)(\alpha  + 2)}}{{4\alpha (5 - 2\omega )(7 - 2\omega )}},
\end{equation}
where the molecular scattering factor $\alpha$ and the heat index $\omega$ depend on the type of gas molecule. The limiting factor ${\mathfrak{q}}(\kappa)$ is constructed to weaken the viscous term in large Kn number case to make the scheme stable. It can be seen from Eq.~\ref{eq:mac_interfacef_nccrlim} and Eq.~\ref{eq:mac_interfacef_nccrkap} that, when the stress and heat flux are small, ${\mathfrak{q}}(\kappa)$ is approaching to $1$ and we can get the NS viscous term in Eq.~\ref{eq:mac_interfacef}, when the stress and heat flux are large, ${\mathfrak{q}}(\kappa)$ is approaching to $0$ and the viscous term is weakened.  Here we further reveal the mechanism of ${\mathfrak{q}}(\kappa)$ through a simple one-dimensional case where there is no stress but only heat flux, i.e.~$k\partial T/\partial x \ne 0$ and $\partial {U}/\partial x = 0$. In this case $\kappa$ is
\begin{equation}
\kappa  = \ln \left( {2\frac{{{\pi ^{\frac{1}{4}}}}}{{\sqrt {2\beta } }}\sqrt {\frac{{\Pr {{\left| {k\partial T/\partial x} \right|}^2}}}{{{C_p}T{p^2}}}}  + 1} \right),
\end{equation}
and the heat flux from the viscous term of Eq.~\ref{eq:mac_interfacef} is
\begin{equation}\label{eq:mac_qflux_nccrlim}
q = {\mathfrak{q}}(\kappa)\left( { - k\frac{{\partial T}}{{\partial x}}} \right) = {\mathfrak{q}}(\kappa){q_{{\rm{NS}}}}.
\end{equation}
If the magnitude of the NS heat flux $\left|{q_{{\rm{NS}}}}\right|$ approaches $0$, ${\mathfrak{q}}(\kappa)$ will approach $1$ and $q$ approaching $q_{{\rm{NS}}}$ holds true for Eq.~\ref{eq:mac_qflux_nccrlim}. If $\left|{q_{{\rm{NS}}}}\right|$ approaches $+\infty$, in this case the heat flux $q$ from Eq.~\ref{eq:mac_qflux_nccrlim} goes to
\begin{equation}\label{eq:mac_qflux_nccrlim_infty}
q = \frac{{\ln \left( {2M\left| {{q_{{\rm{NS}}}}} \right|} \right)}}{M}\frac{{{q_{{\rm{NS}}}}}}{{\left| {{q_{{\rm{NS}}}}} \right|}},
\end{equation}
where $M$ is
\begin{equation}\label{eq:mac_qflux_nccrlim_M}
M = \frac{\pi ^{\frac{1}{4}}}{{\sqrt {2\beta } }}\sqrt {\frac{{\Pr }}{{{C_p}T{p^2}}}}.
\end{equation}
On the other hand, in the NCCR relation \cite{liu2019Anextended}, for one-dimensional case, if $\partial {U}/\partial x = 0$, the heat flux is calculated as
\begin{equation}\label{eq:nccr_qflux_1d}
{q_{{\rm{NCCR}}}} = {\mathfrak{q}}({\kappa _{{\rm{NCCR}}}})\left( { - k\frac{{\partial T}}{{\partial x}}} \right) = {\mathfrak{q}}({\kappa _{{\rm{NCCR}}}}){q_{{\rm{NS}}}},
\end{equation}
where ${\kappa _{{\rm{NCCR}}}}$ is
\begin{equation}\label{eq:nccr_kapa_1d}
{\kappa _{{\rm{NCCR}}}} = \frac{{{\pi ^{\frac{1}{4}}}}}{{\sqrt {2\beta } }}\sqrt {\frac{{\Pr {{\left| {{q_{{\rm{NCCR}}}}} \right|}^2}}}{{{C_p}T{p^2}}}}.
\end{equation}
If the magnitude of $\left|{q_{{\rm{NCCR}}}}\right|$ approaches $0$, similarly $q_{{\rm{NCCR}}}$ approaching $q_{{\rm{NS}}}$ holds true, i.e.~the NS heat flux is recovered. If the magnitude of $\left|{q_{{\rm{NCCR}}}}\right|$ approaches $+\infty$, in this limiting case the magnitude of the heat flux can be deduced from Eq.~\ref{eq:nccr_qflux_1d} and Eq.~\ref{eq:nccr_kapa_1d} as
\begin{equation}\label{eq:nccr_qflux_1d_infty}
\left| {{q_{{\rm{NCCR}}}}} \right| = \frac{{\ln \left( {2M\left| {{q_{{\rm{NS}}}}} \right|} \right)}}{M},
\end{equation}
where $M$ has the same definition as Eq.~\ref{eq:mac_qflux_nccrlim_M}. Comparing Eq.~\ref{eq:mac_qflux_nccrlim_infty} and Eq.~\ref{eq:nccr_qflux_1d_infty}, one can find that $q$ and $q_{{\rm{NCCR}}}$ are identical in the limiting case. The above derivation implies that $q$ and $q_{{\rm{NCCR}}}$ are very similar when their magnitudes are very small or very large. Of course, instead of the above special case, for more general multidimensional case, $\vec q$ from the viscous term of Eq.~\ref{eq:mac_interfacef} and $\vec q_{{\rm{NCCR}}}$ based on the NCCR relation \cite{liu2019Anextended} are not exactly same when their magnitudes approach $+\infty$, but they are generally of the same order of magnitude when they are large. All in all, the viscous term of Eq.~\ref{eq:mac_interfacef} recovers the NS viscous term when the stress and heat flux are small, and this viscous term will be reduced to the same order of magnitude as the NCCR viscous term when the stress and heat flux are large. Thus, in small Kn number case the flux obtained from Eq.~\ref{eq:mac_interfacef} is accurate as the NS flux, while in large Kn number case the viscous term of Eq.~\ref{eq:mac_interfacef} is suppressed to make the numerical system more stable.

Finally, take moments of $u_x\vec \psi$ to Eq.~\ref{eq:mac_interfacef} (ignore second and higher order terms of $\tau$, i.e.~$\int {\vec \psi \left( {\frac{{\partial g}}{{\partial t}} + \vec u \cdot \frac{{\partial g}}{{\partial \vec x}}} \right)d\Xi }  = \vec 0$ is used to transform time derivatives into spatial derivatives), the prediction flux is
\begin{equation}\label{eq:mac_predict_flux}
\begin{aligned}
\vec {\mathcal{F}} = & \frac{{\tau '}}{{\tau ' + h}}\int {{u_x}\left( \begin{aligned}
&1\\
&{u_x}\\
&{u_y}\\
&{u_z}\\
&{\frac{1}{2}}{{\vec u}^2}
\end{aligned} \right){g^{{\rm{lr}}}}d\Xi }  + \frac{h}{{\tau ' + h}}\left( \begin{aligned}
&\rho {U_x}\\
&\rho {U_x}{U_x} + p\\
&\rho {U_y}{U_x}\\
&\rho {U_z}{U_x}\\
&(\rho E + p){U_x}
\end{aligned} \right) \\
& + {\mathfrak{q}}(\kappa )\left( \begin{aligned}
& \;\quad 0\\
& - 2\mu {S_{xx}}\\
& - 2\mu {S_{xy}}\\
& - 2\mu {S_{xz}}\\
& - 2\mu {{\vec S}_x} \cdot \vec U - k\frac{{\partial T}}{{\partial x}}
\end{aligned} \right)
\end{aligned}.
\end{equation}
For the calculation about the moments of the Maxwellian distribution function, one can refer to reference \cite{xu2001gas} for some instruction.

The present prediction solver based on Eq.~\ref{eq:mac_predict_flux} is efficient compared to the solver of GKS \cite{xu2001gas}. It is accurate as an NS solver in the continuum flow regime and has enhanced stability in large Kn number case. It is not accurate for rarefied flow calculation, but as mentioned before, it's unrealistic for a solver based on macroscopic variables to provide a very precise flux in large Kn number case. As a prediction solver, stability is the most important. The accuracy of the final solution obtained from the present method only depends on the microscopic scheme described in Section \ref{sec:loop_l}.

\section{Numerical results and discussions}\label{sec:numericaltest}
More test cases will be added during the preparation of the final paper.

\subsection{Lid-driven cavity flow}\label{sec:testcav}
\begin{figure}
\centering
\subfigure[\label{fig:testcav_meshnonuni}Nonuniform mesh for Re=1000 ]{\includegraphics[width=0.47\textwidth]{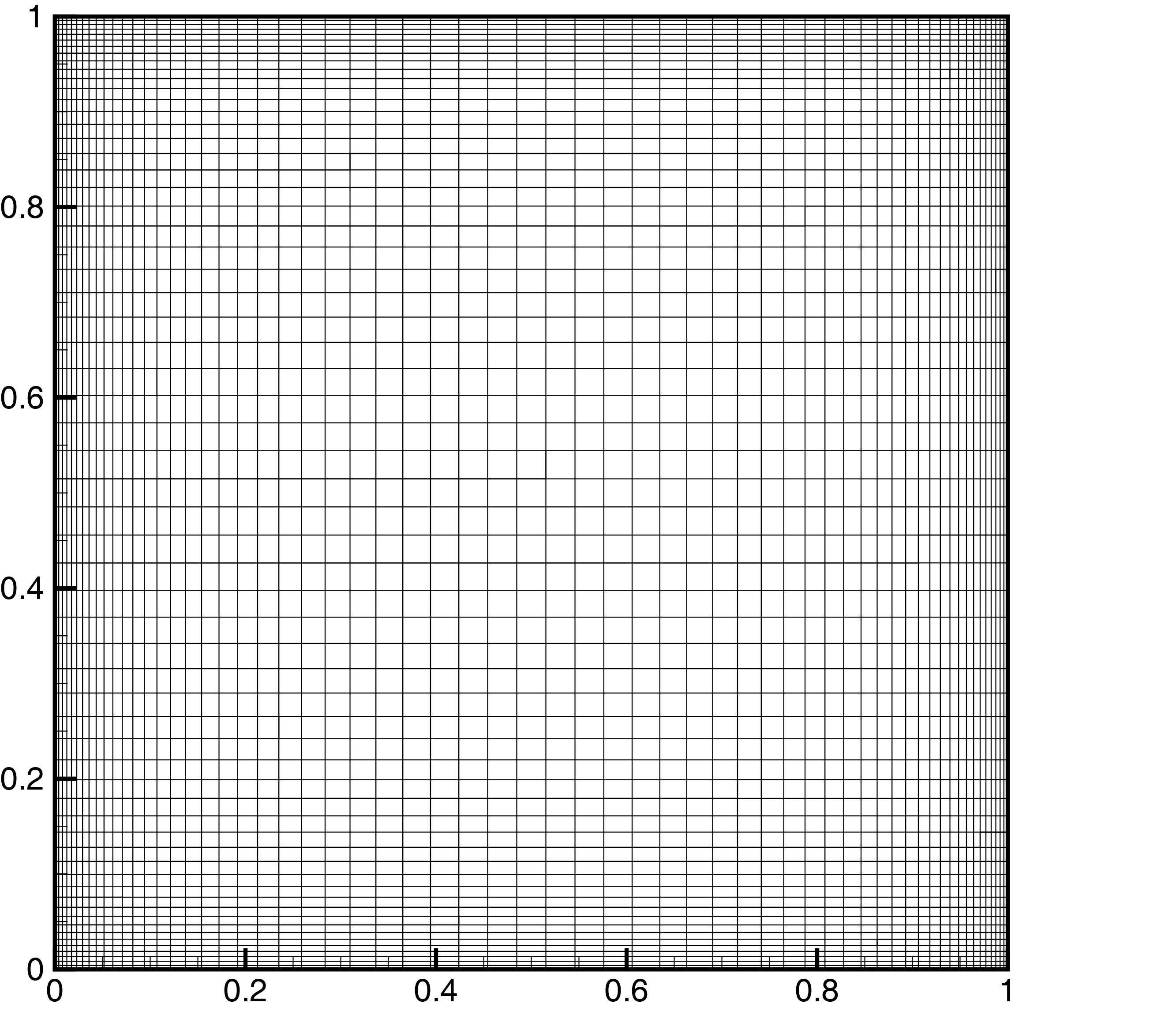}}\hspace{0.02\textwidth}%
\subfigure[Uniform mesh for Kn=0.075, 10]{\includegraphics[width=0.47\textwidth]{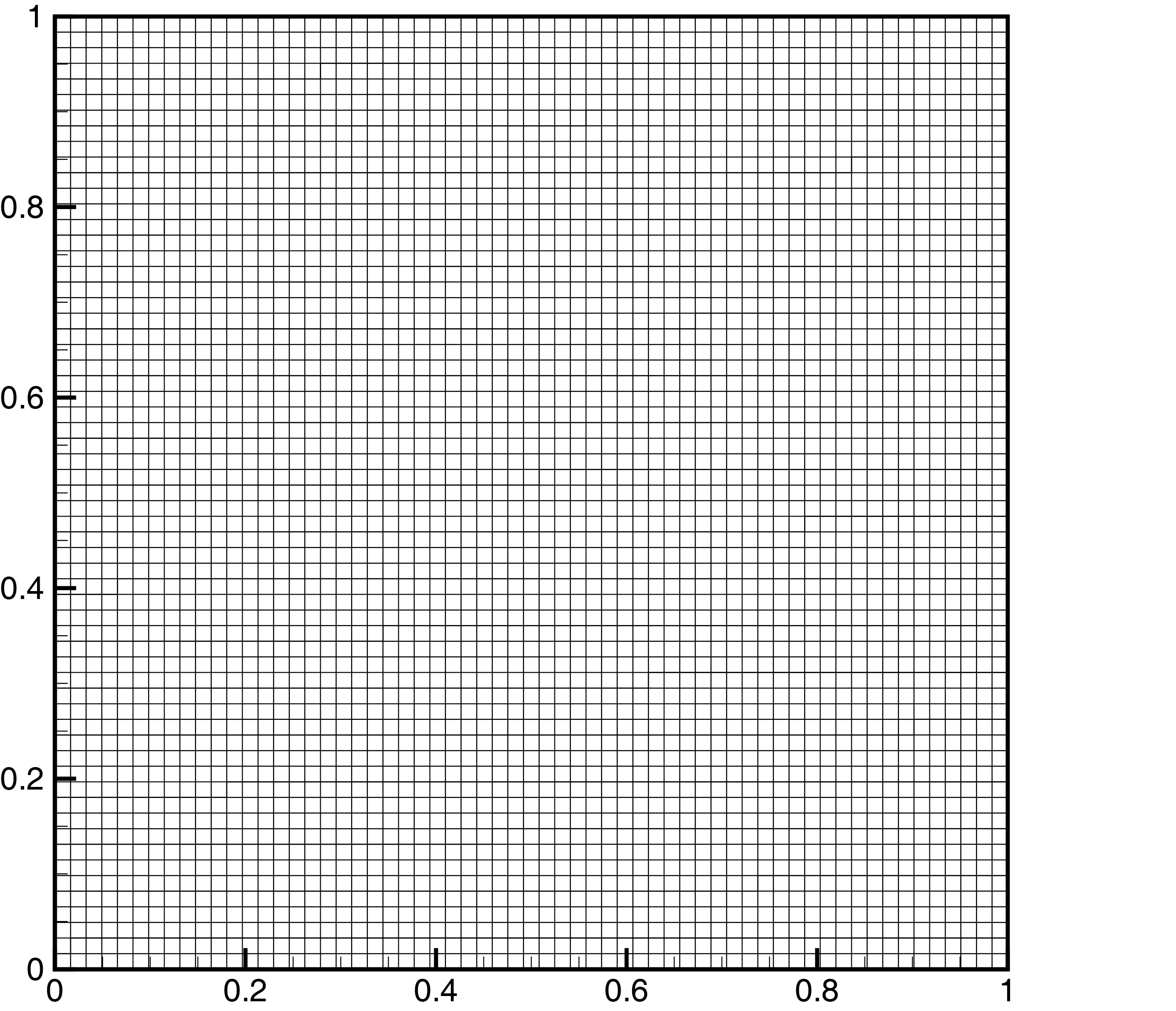}}
\caption{\label{fig:testcav_mesh}Physical space mesh (61$\times$61) for cavity flow simulations.}
\end{figure}

\begin{figure}
\centering
\includegraphics[width=0.47\textwidth]{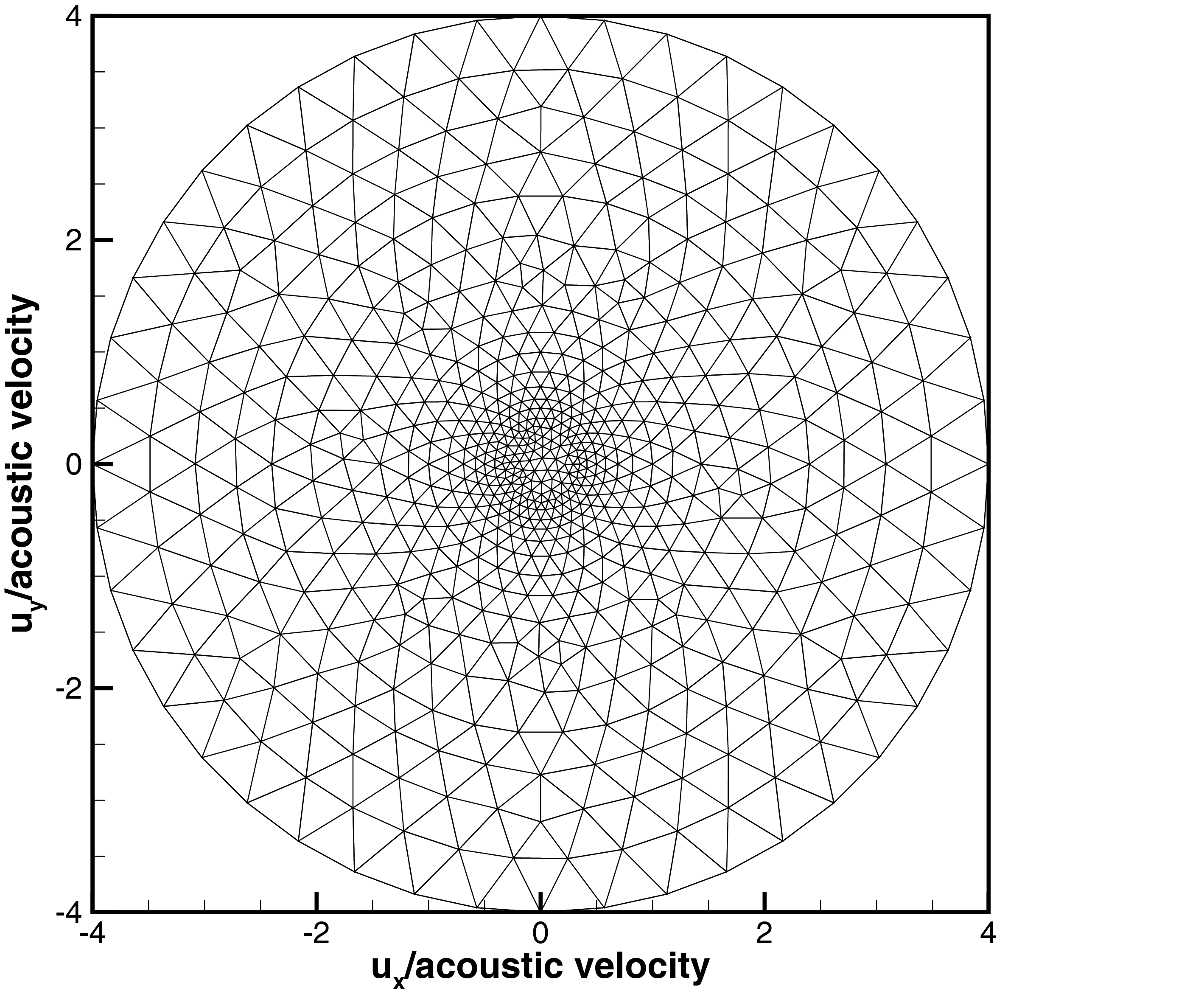}
\caption{\label{fig:testcav_meshmic}Velocity space mesh (1192 cells) for cavity flow simulations.}
\end{figure}

\begin{figure}
\centering
\subfigure[Streamlines]{\includegraphics[width=0.47\textwidth]{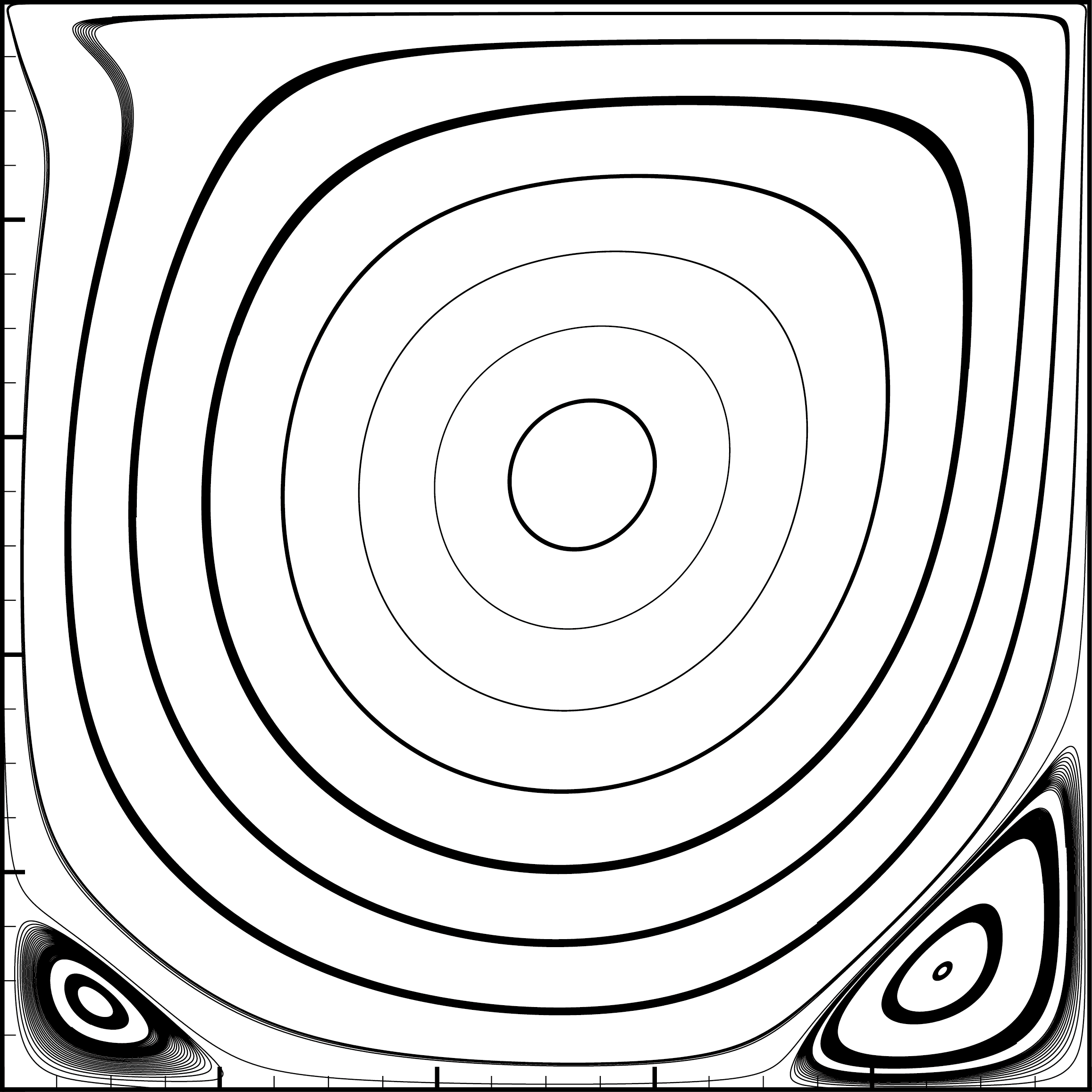}}\hspace{0.02\textwidth}%
\subfigure[$U_y$ along the horizontal central line and $U_x$ along the vertical central line]{\includegraphics[width=0.47\textwidth]{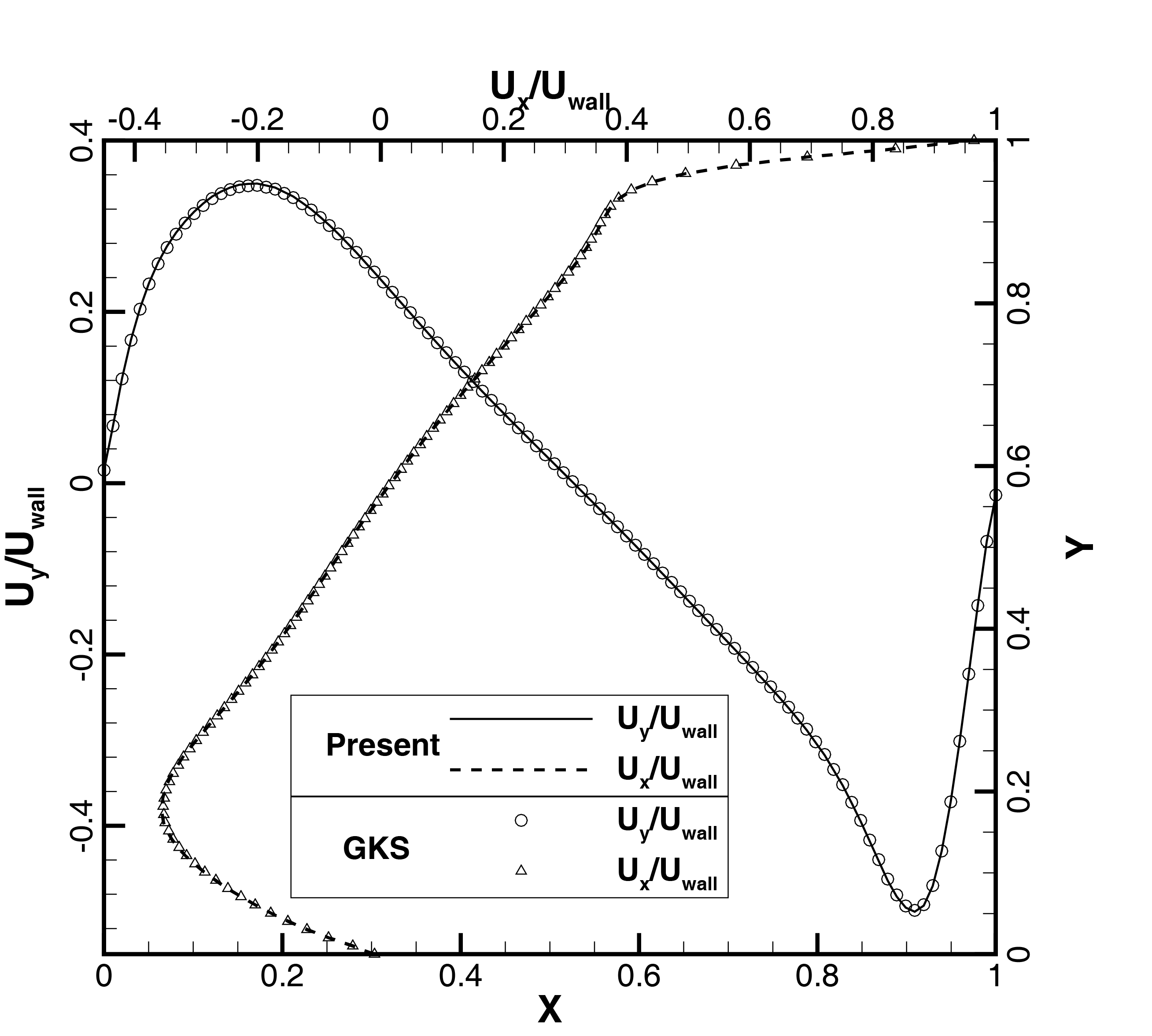}}
\caption{\label{fig:testcav_1000}Cavity flow at Re=1000. The reference result is calculated by GKS \cite{xu2001gas} without discretization of velocity space (identical to Navier-Stokes solution).}
\end{figure}

\begin{figure}
\centering
\subfigure[Temperature contours, color band: UGKS, dashed line: present]{\includegraphics[width=0.47\textwidth]{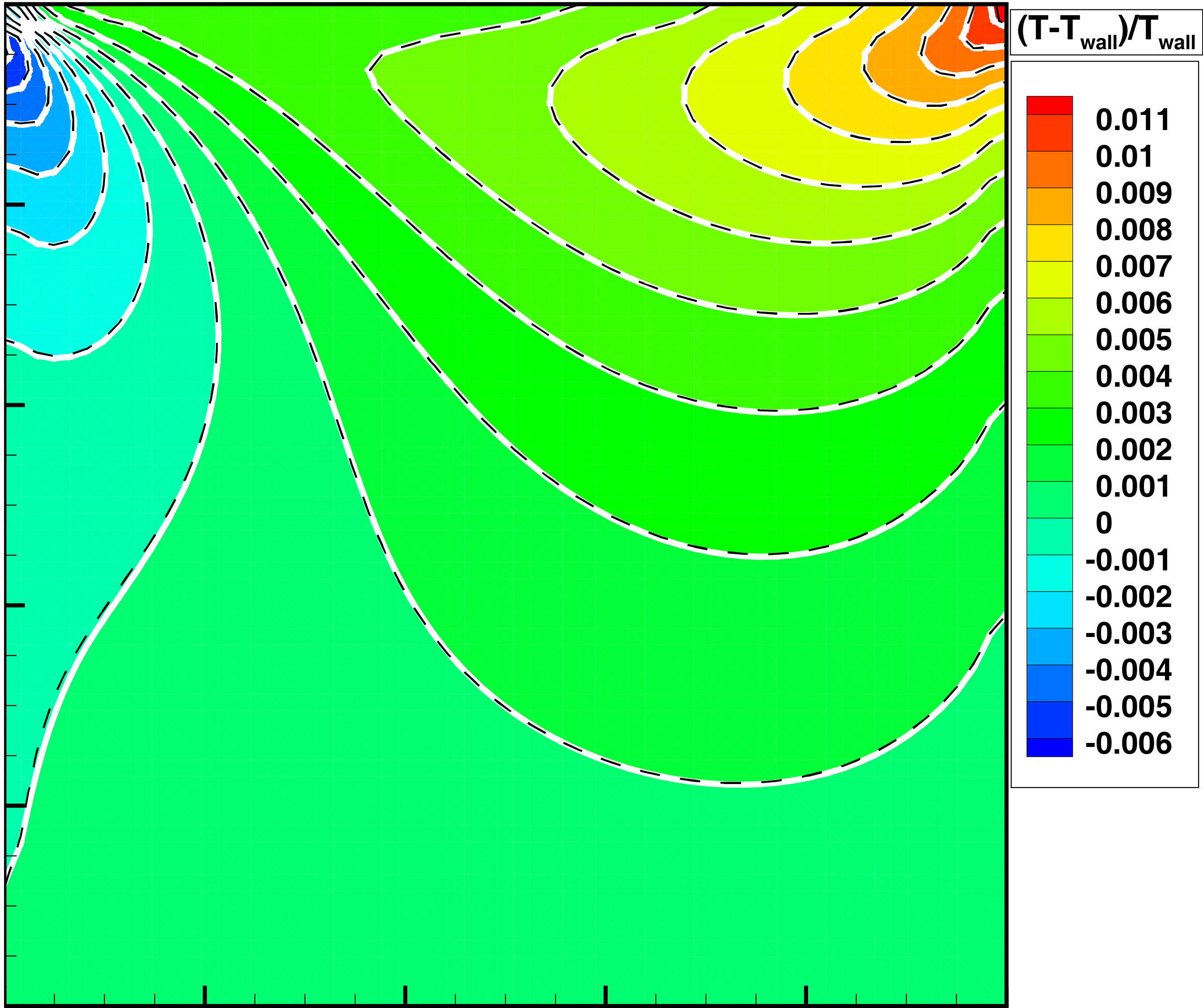}}\hspace{0.02\textwidth}%
\subfigure[Heat flux, circle: UGKS, line: present]{\includegraphics[width=0.47\textwidth]{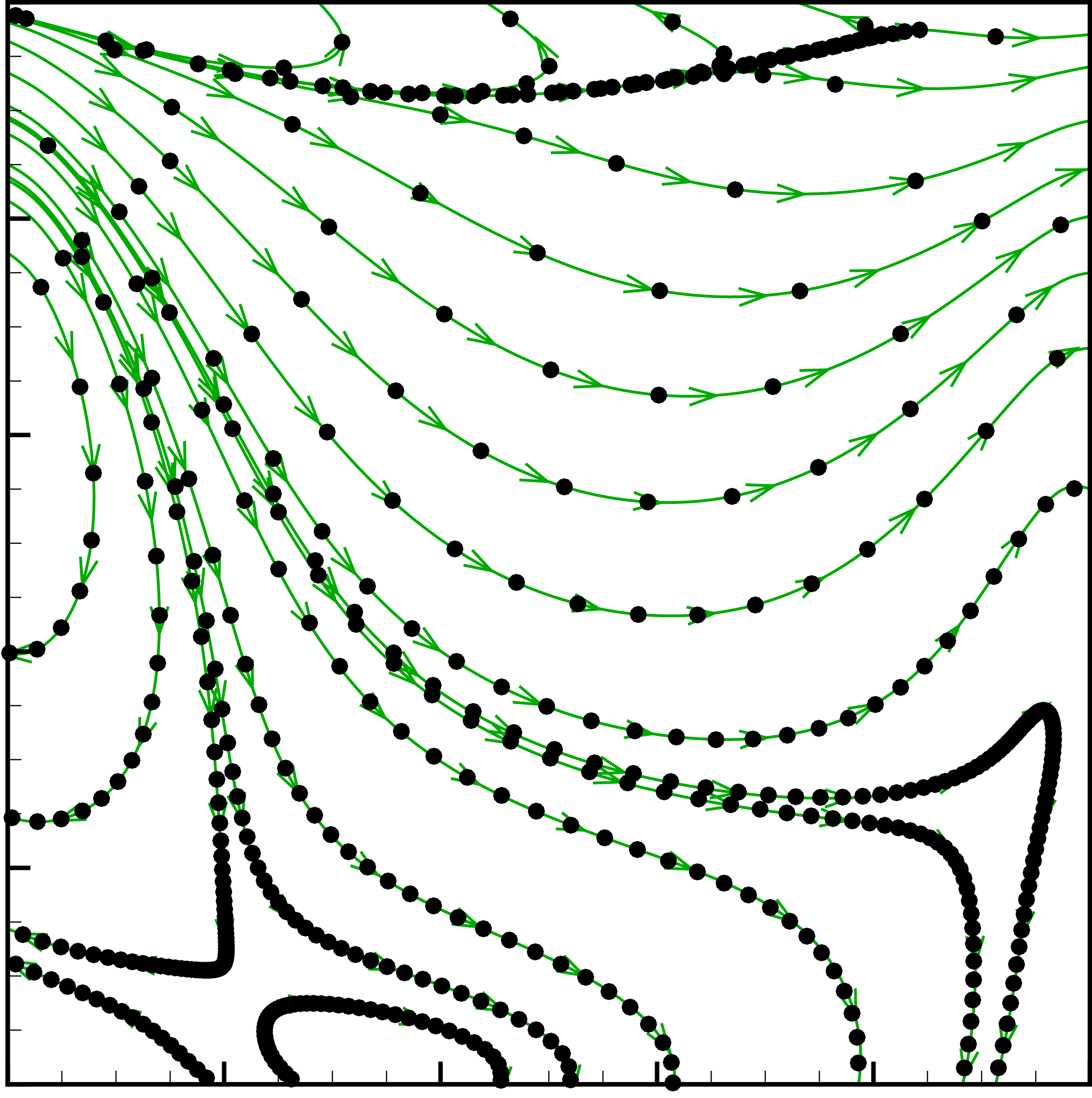}}\\
\subfigure[Streamlines]{\includegraphics[width=0.47\textwidth]{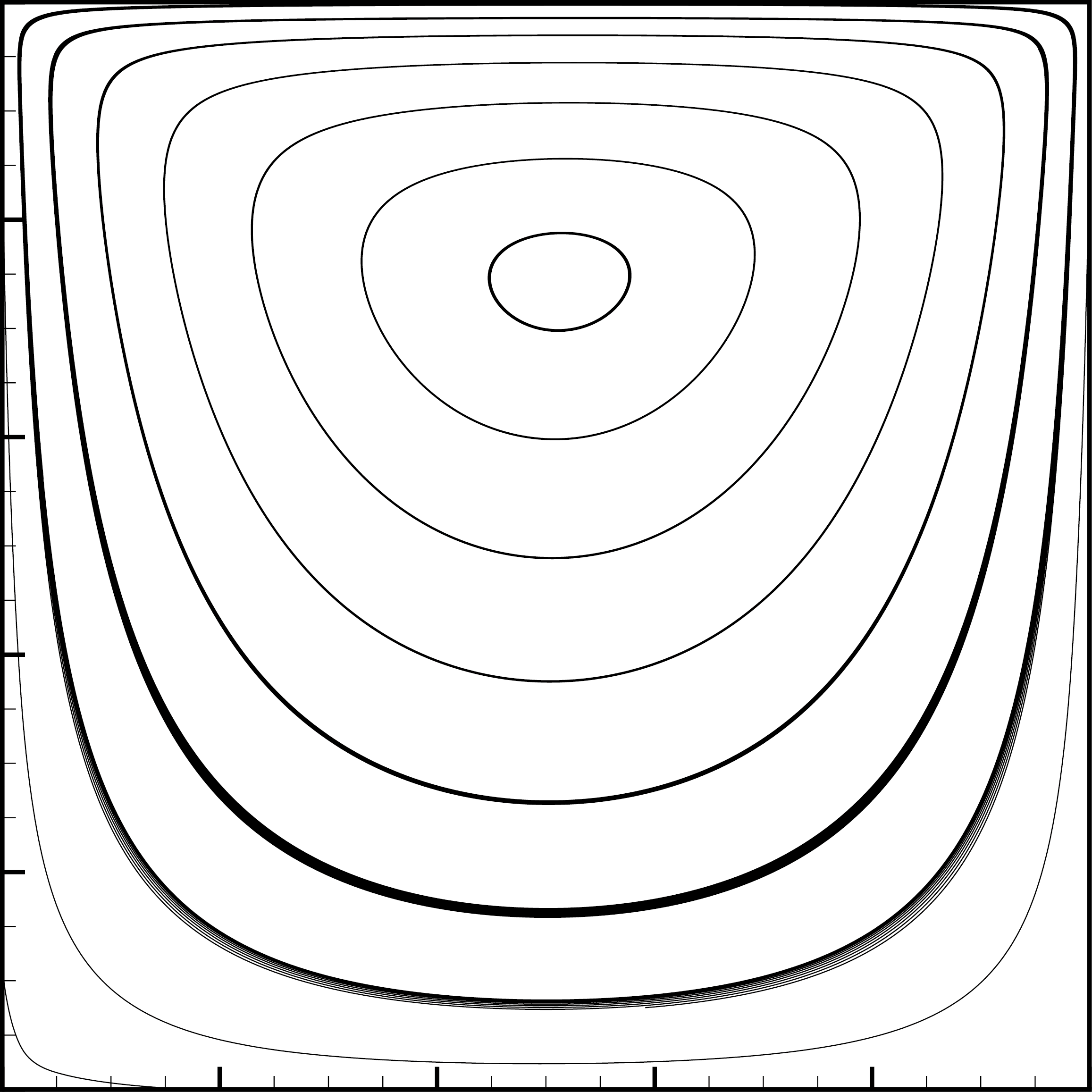}}\hspace{0.02\textwidth}%
\subfigure[$U_y$ along the horizontal central line and $U_x$ along the vertical central line]{\includegraphics[width=0.47\textwidth]{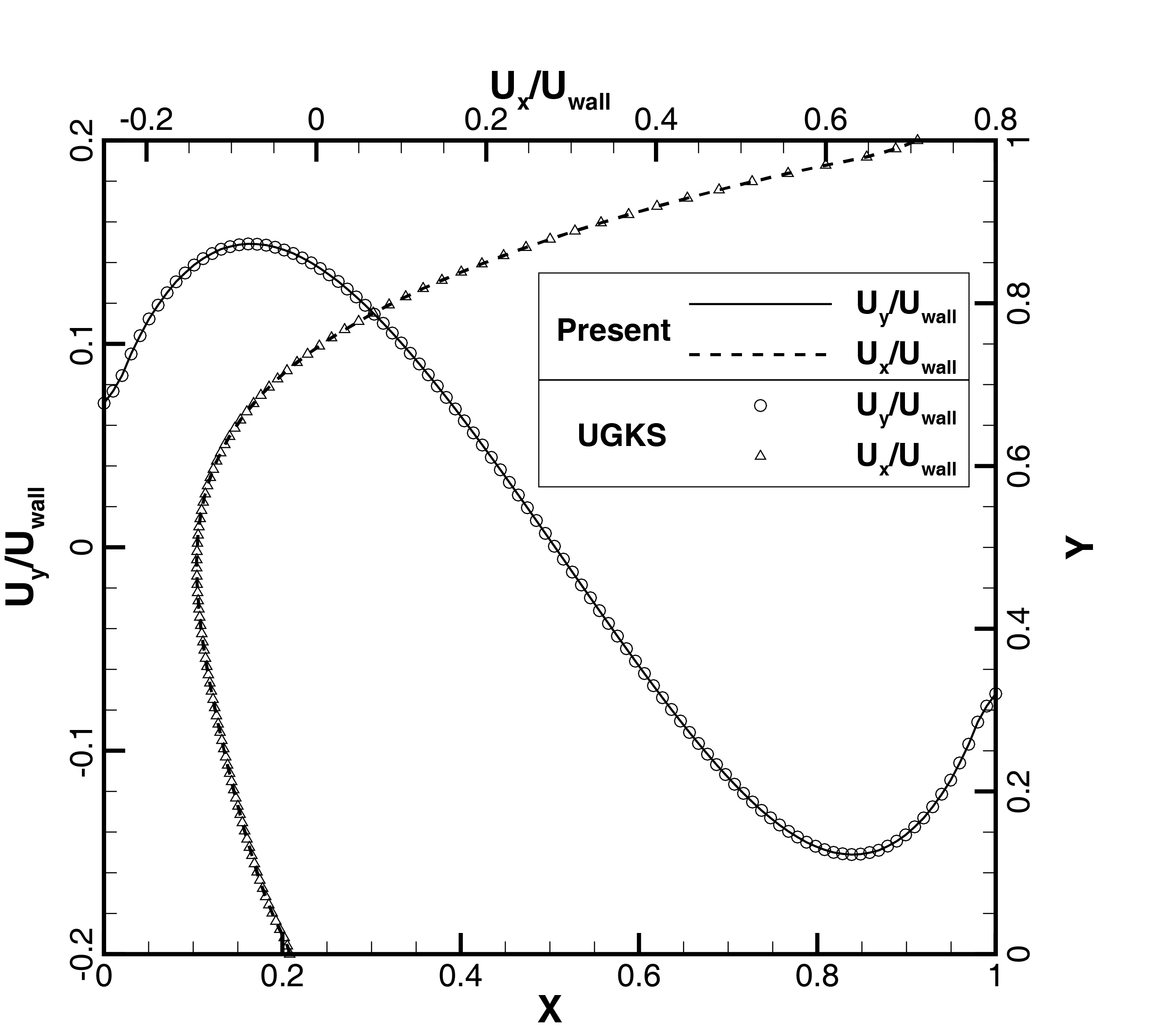}}
\caption{\label{fig:testcav_75}Cavity flow at Kn=0.075. The reference result is calculated by UGKS \cite{Xu2010A}.}
\end{figure}

\begin{figure}
\centering
\subfigure[Temperature contours, color band: UGKS, dashed line: present]{\includegraphics[width=0.47\textwidth]{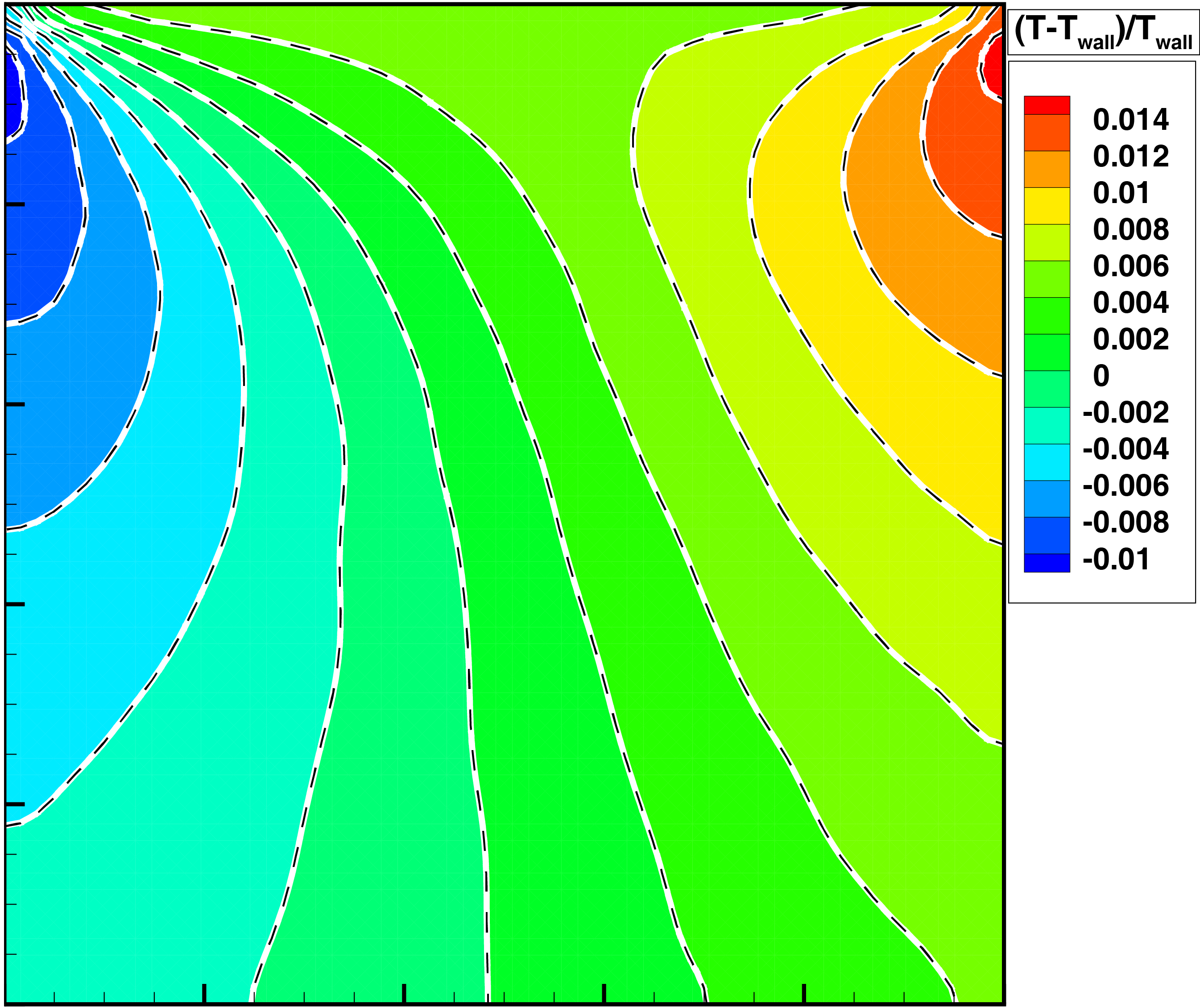}}\hspace{0.02\textwidth}%
\subfigure[Heat flux, circle: UGKS, line: present]{\includegraphics[width=0.47\textwidth]{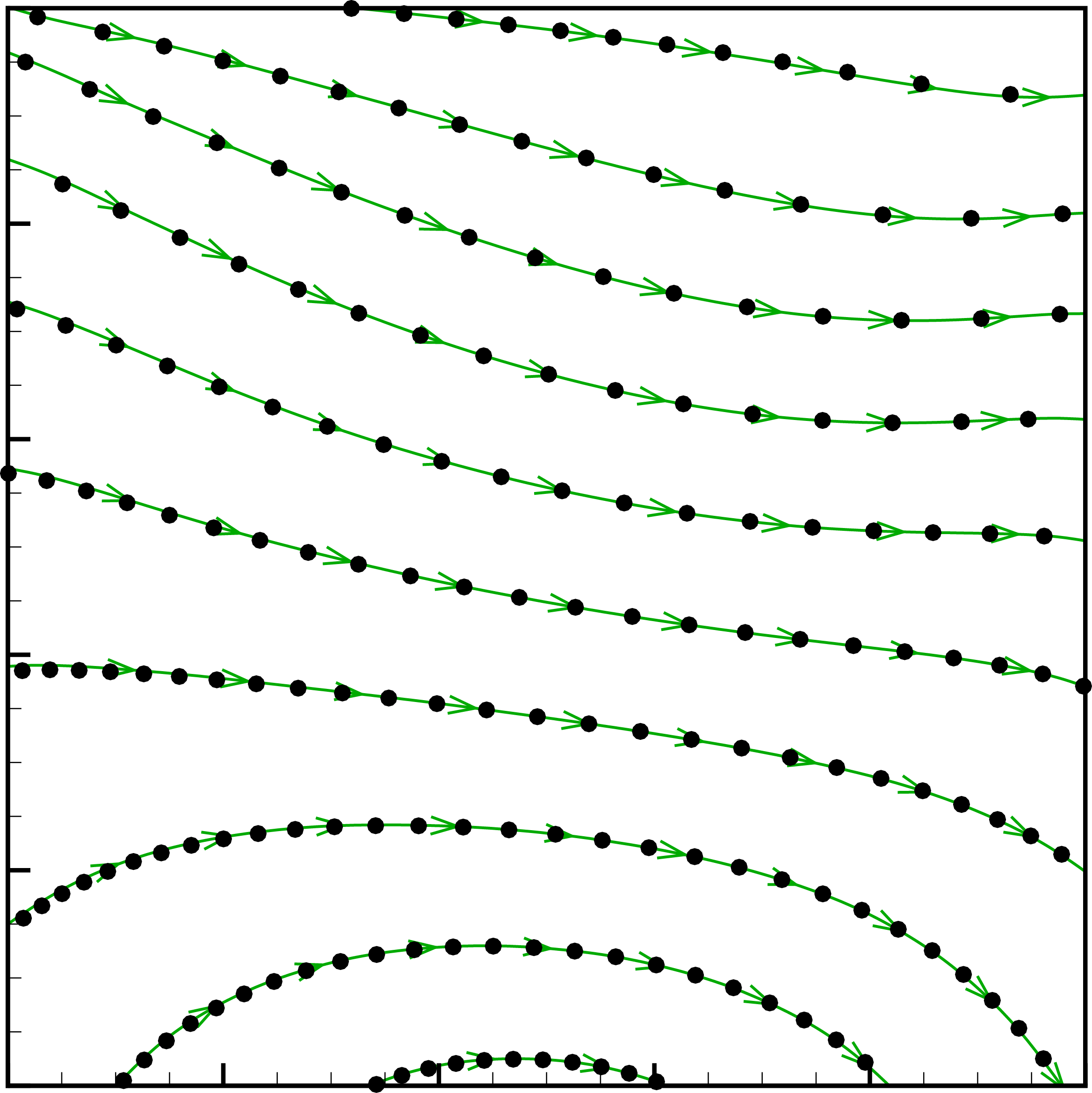}}\\
\subfigure[Streamlines]{\includegraphics[width=0.47\textwidth]{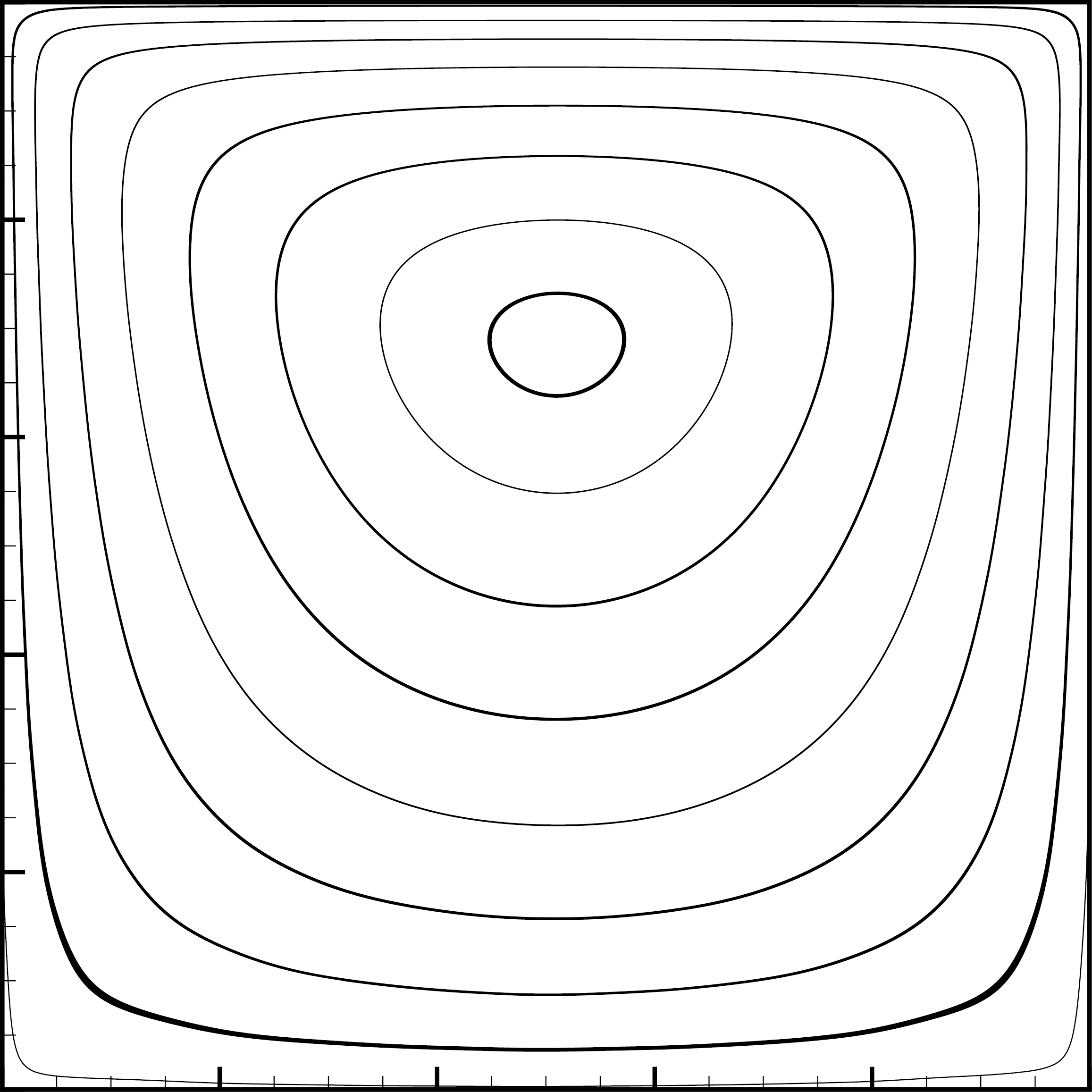}}\hspace{0.02\textwidth}%
\subfigure[$U_y$ along the horizontal central line and $U_x$ along the vertical central line]{\includegraphics[width=0.47\textwidth]{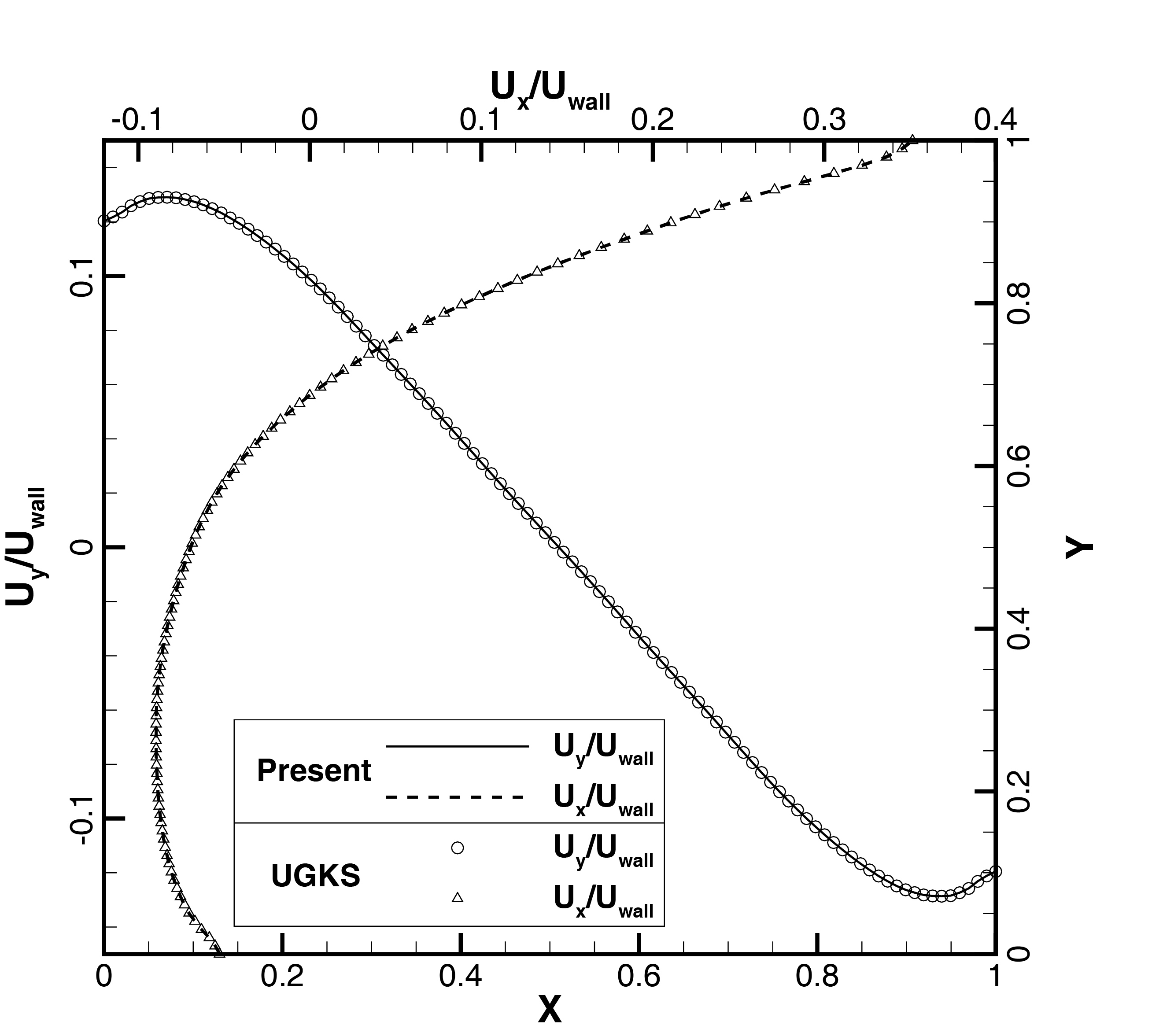}}
\caption{\label{fig:testcav_10}Cavity flow at Kn=10. The reference result is calculated by UGKS \cite{Xu2010A}.}
\end{figure}

The test case of lid-driven cavity flow is performed to test the efficiency of the present method, and to test if the viscous effect can be correctly simulated by the present method. Three cases Re=1000 and Kn=0.075, 10 are considered, involving gas flows from continuum regime to free molecular regime. The Mach number, which is defined by the upper wall velocity $U_{\rm{wall}}$ and the acoustic velocity, is 0.16. The Shakhov model is used and the Prandtl number Pr=$2/3$. The hard sphere (HS) model is used, with heat index $\omega$=0.5 and molecular scattering factor $\alpha$=1. On the wall of the cavity, the diffuse reflection boundary condition with full thermal accommodation \cite{Xu2015Direct} is implemented. For the physical space discretization, as shown in Fig.~\ref{fig:testcav_mesh}, a nonuniform 61$\times$61 mesh with a mesh size $0.004L$ ($L$ is the width of the cavity) near the wall is used for the case Re=1000 while a uniform 61$\times$61 mesh is used for the cases Kn=0.075, 10. For the velocity space discretization, as shown in Fig.~\ref{fig:testcav_meshmic}, a 1192 cells' unstructured mesh is used, where the central area is refined to reduce the ray effect. For the iteration strategy, in each step $n$, 60 turns of $m$ loop and 3 turns of $l$ loop are performed, while 40 times and 6 times of SGS iterations are executed for each turn of the $m$ loop and the $l$ loop respectively. The prediction step $\Delta t_i^{n + 1}$ in Eq.~\ref{eq:mac_prediction} is set as $ + \infty $ in this set of test cases. The convergence criterion is that the global root-mean-square of the infinite norm about the macroscopic residual vector defined by Eq.~\ref{eq:mac_fluxrsd} is less than $10^{-9}$. Computations are run on a single core of a computer with \emph{Intel(R) Xeon(R) CPU E5-2678 v3 @ 2.50GHz}. The computational efficiency compared with the implicit multiscale method in reference \cite{yuan2018conservative} is shown in Tab.~\ref{tab:testcav_eff}. It can be seen that in the continuum flow regime (case of Re=1000), the present method is one order of magnitude faster than the implicit method of reference \cite{yuan2018conservative}. Considering that the implicit method of reference \cite{yuan2018conservative} is two orders of magnitude faster than explicit UGKS (discussed in \cite{yuan2018conservative}) in the continuum flow regime, the present method should be thousands of times faster than explicit UGKS in the continuum flow regime. For the cases Kn=0.075, 10, the present method is only one to two times faster than the method of reference \cite{yuan2018conservative}. This efficiency is reasonable because in the continuum flow regime the prediction scheme (the $m$ loop) gives very accurate predicted macroscopic variables and the numerical system converges rapidly, while for the rarefied flow the prediction scheme fails to be so precise and the slight efficiency increase compared with the method of reference \cite{yuan2018conservative} is due to the improved iteration strategy (namely, the $l$ loop) of the present method. The results of the present method for this set of test cases are shown in Fig.~\ref{fig:testcav_1000}, Fig.~\ref{fig:testcav_75} and Fig.~\ref{fig:testcav_10}. The present results agree very well with the results of GKS and UGKS.

\clearpage


\bibliographystyle{yuan_mimplicit}
\bibliography{yuan_mimplicit}

\clearpage

\clearpage

\clearpage
\renewcommand{\multirowsetup}{\centering}

\begin{table}
\centering
\caption{\label{tab:testcav_eff}Comparison of the efficiency between the implicit method in reference \cite{yuan2018conservative} and the present method for cavity flow simulations.}
\begin{tabular}{p{50pt} p{40pt}<{\raggedleft} p{40pt}<{\raggedleft} p{40pt}<{\raggedleft} p{40pt}<{\raggedleft} p{40pt}<{\raggedleft} p{30pt}<{\raggedleft}}
\hline

\hline
\multicolumn{1}{c}{\multirow{2}{*}{Case}} & \multirow{2}{40pt}{Velocity space} & \multicolumn{2}{c}{Implicit method \cite{yuan2018conservative}} & \multicolumn{2}{c}{Present} & \multicolumn{1}{c}{\multirow{2}{*}{Speedup}}\\\cline{3-4}\cline{5-6}
 ~&~& \multicolumn{1}{c}{Steps} & \multicolumn{1}{c}{Time (s)} & \multicolumn{1}{c}{Steps} & \multicolumn{1}{c}{Time (s)} &~ \\
\hline
Re=1000 & 1192 & 865 & 1102 & 23 & 74.4 & 14.8 \\
Kn=0.075 & 1192 & 117 & 150 & 28 & 93.3 & 1.6 \\
Kn=10 & 1192 & 173 & 225 & 33 & 114.3 & 2.0 \\
\hline

\hline
\end{tabular}
\end{table}

\end{document}